\documentclass[journal=jacsat,manuscript=article]{achemso}

\usepackage{multirow}
\usepackage{setspace}
\usepackage[version=3]{mhchem} 
\usepackage{xcolor}
\usepackage{soul}
\usepackage{chemformula} 
\usepackage[colorlinks=true,allcolors=blue]{hyperref}

\author{Harikrishnan G.}
\affiliation[Unknown University]
{School of Physics, Indian Institute of Science Education and Research Thiruvananthapuram, Kerala 695551, India}
\author{Shashwata Chattopadhyay}
\affiliation[Unknown University]
{School of Physics, Indian Institute of Science Education and Research Thiruvananthapuram, Kerala 695551, India}
\author{Arijit Kayal}
\affiliation[Unknown University]
{School of Physics, Indian Institute of Science Education and Research Thiruvananthapuram, Kerala 695551, India}
\author{K. Bandopadhyay}
\affiliation[Second university]
{Institute of Microelectronics and Photonics, University of Warsaw, Poland}
\author{K. Kolodziejak}
\affiliation[Second university]
{Institute of Microelectronics and Photonics, University of Warsaw, Poland}
\author{Vinayak B. Kamble}
\affiliation[Unknown University]
{School of Physics, Indian Institute of Science Education and Research Thiruvananthapuram, Kerala 695551, India}
\author{Dorota A. Pawlak}
\affiliation[Second university]
{Institute of Microelectronics and Photonics, University of Warsaw, Poland}
\alsoaffiliation[Second University]
{$Ensemble^3$ Centre of Excellence, Warsaw, Poland}
\author{J. Mitra}
\affiliation[Unknown University]
{School of Physics, Indian Institute of Science Education and Research Thiruvananthapuram, Kerala 695551, India}
\email{j.mitra@iisertvm.ac.in}

\title
{Photo-thermoelectrics of a distributed Schottky junction system: Delineating the photo and thermal currents}
\abbreviations{SEM, XRD, XPS, UV}
\SectionNumbersOn
\begin{document}
	
		
		
		
		
		
	
	\begin{abstract}
	Hybrid materials, consisting of diverse materials combined in various configurations, exhibit intriguing opto-electronic properties and are a focal point of research in functional materials. However, achieving precise predictability and tunability of their macroscopic properties in terms of the control parameters remains a challenge. One promising approach involves leveraging self-organization through eutectic growth and subsequent engineering via annealing to engineer niche properties. Here, the electronic properties of eutectic \ch{NiTiO3}–\ch{TiO2} samples and their \ch{H2}-reduced counterpart \ch{Ni-TiO2} are investigated, where the latter feature high aspect ratio \ch{TiO2} nanostructures decorated with nodular \ch{Ni} globules. By exploiting this unique  architecture and capitalizing on the nanostructuring processes alongside material properties, performance  enhancement of photoactive devices is shown. Further, the competing mechanisms of photo-driven and photo-thermal-driven transport is delineated to characterize the overall photo-response of the system. The ability for self-powered functionality further showcases this approach as a promising strategy for developing efficient self-powered devices.
	\end{abstract}

	\section{Introduction}
At the core of engineering efficient devices lies selection of appropriate materials, combined in a conducive form factor for optimal performance. Today, hybrid materials stand at the forefront in materials engineering, blending metals and semiconductors, organics and inorganics, polymers and ceramics to harness their synergistic responses and unlock novel functionalities. By integrating materials with contrasting properties at the nanoscale or molecular level, these complex systems exhibit  properties that surpass those of their individual components or nucleate entirely new functionalities absent in the components. Such materials have found applications in biomedicine, power generation and storage, as well as in optoelectronic and photo-catalytic technologies, among others. \cite{saveleva2019hierarchy,gu2018introducing,nicole2014hybrid,gibson2010review}. 
	These novelties stem from the contrasting material properties, especially of the interfaces. The interfaces  play a crucial role in realizing functionalities unattainable in homogeneous systems, e.g. rectification, switching, waveguides etc. While initial hybrid material fabrication were limited to liquid state mixes or planar multi-layers, advancements in nanoengineering and materials processing capabilities today allow complex three-dimensional architectures that has revolutionized the design and fabrication landscape.  
	
	Composite materials are typically prepared by both bottom-up and top-down techniques. Further, the use of eutectic transitions in synthesizing composites result in highly crystalline biphasic materials with controllable microstructure dimensions and pristine interfaces, via a one-step process. Though, eutectic composites have been investigated mainly due to their favourable mechanical properties, recently they are being used in thermoelectric and thermophotovoltaic systems \cite{snyder2008complex,sola2019directionally},  photoelectrochemistry \cite{kolodziejak2023durability}, optical metamaterials \cite{osewski2020new,kim2015template,petronijevic2023surprising}, and as active plasmonic materials \cite{sadecka2015eutectics,szlachetko2020selective}. These eutectic composites exhibit exceptional potential for fabricating a wide variety of self-organized functional materials.
	Here, an eutectic based bottom-up route  is employed, with post-processing, to engineer a nanostructured hybrid with distributed metal-semiconductor interfaces. 
	Beginning with a phase segregated \ch{NiTiO3}-\ch{TiO2} (NTT) eutectic, prepared by the micro-pulling-down ($\mu$-PD) method, the final material  is synthesized by high temperature \ch{H2} reduction of NTT to generate a Magn$\acute{\mathrm{e}}$li phase \ch{TiO2} matrix (m\ch{TiO2})\cite{andersson1957phase}  decorated with Ni nanoparticles (NP).
	Semiconductor eutectics are multi-phase systems, amenable to post processing via etching, annealing etc. that often exhibit ordered micro or nanostructures \cite{jackson1988lamellar, pawlak2006self, kulkarni2020archimedean, kolodziejak2016synthesis, osewski2020new}.  
	While their phase segregated nature induce niche mechanical properties \cite{waku1998new}, the high crystallinity of the phases and the atomically sharp interfaces allow for high carrier mobility making them conducive for electronic and photo-stimulated applications \cite{wysmulek2017srtio3, zeeshan2022semiconducting}. 
	Both \ch{NiTiO3} and \ch{TiO2} are wide band gap ($E_g$) semiconductors  (SI fig. S21) with diverse applications \cite{iancu2015atomic, panepinto2021switching, anitha2015recent, ziental2020titanium, noman2019synthesis, li2020impact,kumar2021effect,pintor2019controlling,irwin2008p}, e.g. the Honda-Fujishima effect  established the photo-catalytic supremacy of \ch{TiO2}\cite{fujishima1972electrochemical}. 
 \ch{TiO2}'s formation thermodynamics allows a wide range of  m\ch{TiO2}  (Ti$_n$O$_{2n-1}$)  phases and sub-oxides characterized by band gaps narrower than that of \ch{TiO2} ($E_g^{\ch{TiO2}}\sim$ 3.3 eV) accompanied by higher electrical conductivity \cite{nowotny2008titanium, naldoni2012effect,bartholomew1969electrical, walsh2010continuing}. \ch{NiTiO3} is an ilmenite semiconductor with $E_g^{\ch{NiTiO3}} \sim$ 2.18 eV that has been investigated  in heterostructure assemblies  with \ch{TiO2} to promote charge-separation \cite{li2018heterostructured, xing2019porous}, hole-transportation \cite{wu2017simultaneous} and photocatalytic activity \cite{qu2012facile, inceesungvorn2014novel}.
And Ni-\ch{TiO2} composites have been explored for photo-stimulated applications leveraging the plasmonic response of Ni NPs and interfacial charge segregation  \cite{domaschke2019magneli}. 

Reduction of the NTT eutectic leaches out  Ni that agglomerates as NPs to decorate the  highly porous, dendritic m\ch{TiO2} matrix. Functionalizing these porous systems for photo-stimulated applications would benefit from the elongated dendritic structure of m\ch{TiO2} that aids carrier lifetime and transport via dimensional confinement. The uniformly distributed Ni NP system creates a semiconducting transport network with distributed Schottky interfaces that enable leverage of the band alignment energetics.  Futuristically, the porous architecture also allows fluid diffusion through the bulk that may be exploited for gas sensing applications. Importantly, the sample stoichiometry and its architecture are readily tailored by the parameters associated with the $\mu$-PD technique and subsequent annealing conditions.

This report presents a comprehensive study of the bulk photovoltaic properties of the  Ni NP decorated m\ch{TiO2} system, reduced from the eutectic sample NTT, at  1050 $^\circ$C  as NTT1050 and an intermediate sample NTT700, reduced at 700 $^\circ$C. 
Electrical transport  across 	NTT1050 shows high electrical conductivity, comparable to amorphous carbon \cite{morgan1971electrical,tomidokoro2021electrical} along with 
appreciable photoelectric response, arising from the  distributed Schottky junction network with power generation capability $\sim$ 30 nW per mm$^2$ of illumination. A combination of optoelectronic, thermoelectric and photothermoelectric  investigations showcase the applicability of such systems in realizing self powered devices.  Overall, the investigations yield a comprehensive picture of the hybrid materials comprehending the relevance of the stoichiometry and morphology of the samples and their interplay that determine the overall response.
	
	\begin{figure}
		\begin{center}
			\includegraphics[width=0.95\textwidth]{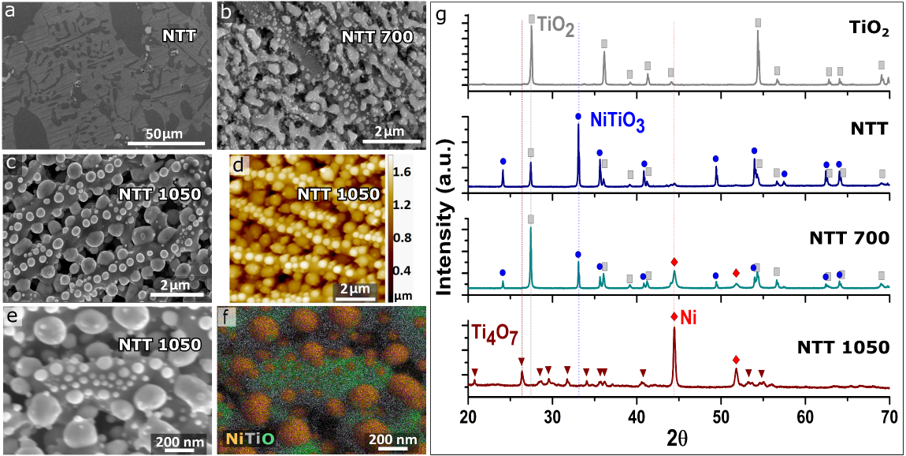}
		\end{center}
		\caption{Characterization of as prepared NiTiO$_3$-TiO$_2$ (NTT) eutectic and H$_2$ annealed (NTT700 and NTT1050) hybrid system. SEM images of (a) NTT, (b) NTT700, (c) NTT1050, (d) AFM topography of NTT1050, (e) SEM image of NTT1050 and corresponding (f) overlaid elemental maps for Ti, O, Ni  and (g) powder XRD data of corresponding samples.}
		\label{fig:SEM-XRD}
	\end{figure}
	
	\section{Results and Discussion}
	\subsection*{Morphology and Stoichiometry}
	The distinct morphologies of the NTT, NTT700 and NTT1050 samples are evident from their secondary electron (SE) images shown in figs. \ref{fig:SEM-XRD}(a-c). The cylindrical samples of diameter $\sim$ 3 mm and thickness $<$ 1 mm are optically polished prior to characterization and device fabrication, (see Supporting Information (SI) section S1 for sample details). The NTT eutectic is phase segregated in the SE image (fig. \ref{fig:SEM-XRD}a) with the lighter regions  being \ch{NiTiO3} and darker being \ch{TiO2}, which are evidenced with greater clarity in the back-scattered electron image shown in fig. S2 and in the optical image in fig. S3d. SI fig. S3 provides the elemental composition of the two phases. The size and fractional composition of the phases are controlled by the $\mu$-PD growth conditions like pulling rate, environment and temperature \cite{binh2010microstructure}.
	The H$_2$ annealed samples NTT700 and NTT1050 evidence the Ni globules that gain clarity and definition upon annealing at the higher $T$ as shown in figs. \ref{fig:SEM-XRD}(b-c).
	The microstructure of NTT1050 with Ni NPs decorating a dendritic network is further visualized in the AFM topography (fig. \ref{fig:SEM-XRD}d), with Ni NP size distribution between 50 nm - 1 $\mu$m. The porous architecture permeates the bulk of the samples as shown in the SE image along cracks and broken edges in (fig. S7).  Fig. \ref{fig:SEM-XRD}(e-f)
	shows overlaid elemental maps for Ni, Ti and O confirming the dominant constituents of the structures.
	Fig. \ref{fig:SEM-XRD}g shows the XRD data of the samples, evidencing NTT as a mix of \ch{NiTiO3} and rutile \ch{TiO2}.  
	The reduced samples show increasing prominence of Ni(111) and Ni(200) crystallites with an initial transition from \ch{NiTiO3} to \ch{TiO2} for NTT700 and later appearance of the m\ch{TiO2} in NTT1050. While the XRD for NTT1050, shown in fig. \ref{fig:SEM-XRD}g, is dominated by the \ch{Ti4O7} phase, fingerprints of other Magn$\acute{\mathrm{e}}$li phases are evidenced across samples as shown in fig. S5.   
	Further evidence of coexisting Magn$\acute{\mathrm{e}}$li phases in NTT1050 is obtained from x-ray photoelectron spectroscopy spectrum for Ti in fig. S8b, showing multiple oxidation states. The Ti$^{4+}$ peak is not purely indicative of \ch{TiO2}, which has negligible signature in the XRD spectrum.  
	Rather phases like \ch{Ti4O7}  support mix-valent  Ti$^{4+}$ and Ti$^{3+}$ states as seen in fig. S8b. 
Presence of the defect related peak in the O1s spectrum (fig. S8d) attest to the presence of O$^{2+}$ in a non-stoichiometric oxygen deficient  environment indicative of  sub-oxides or Magn$\acute{\mathrm{e}}$li phases. 
Deconvolution of the Ni 2p$\frac{3}{2}$ and 2p$\frac{1}{2}$ peaks (fig. S8b) with known satellite structures evidence metallic Ni along with Ni$^{2+}$ states likely originating from surface oxidation of the Ni NPs. 
Optically all three samples appear blackish grey in color with the reflectance spectra shown in fig. S9. Unlike \ch{TiO2}, which shows low absorption in the visible \cite{lee2020application,viswanathan2012nitrogen}, these samples show significant absorption ($>30\%$)  indicative of  oxygen vacancy related sub-band gap states. These are akin to the presence  of black-\ch{TiO2} and Magn$\acute{\mathrm{e}}$li phases that also impart electrical conductivity to the samples.  
While the most reduced sample NTT1050 shows 4-probe electrical conductivity, $\sigma \sim 5\times10^3$ S/m and electron density $n_e \sim$ 10$^{19}$ cm$^{-3}$ at 300 K, the NTT700 and NTT samples show $\sim$ 50 and $\sim$ 1000 times lower conductivity.  NTT1050 exhibits activated transport ($R-T$), attesting to its semiconducting nature as discussed later.  

\subsection*{Optoelectronic properties }
\begin{figure}
		\begin{center}
			\includegraphics[width=0.89\textwidth]{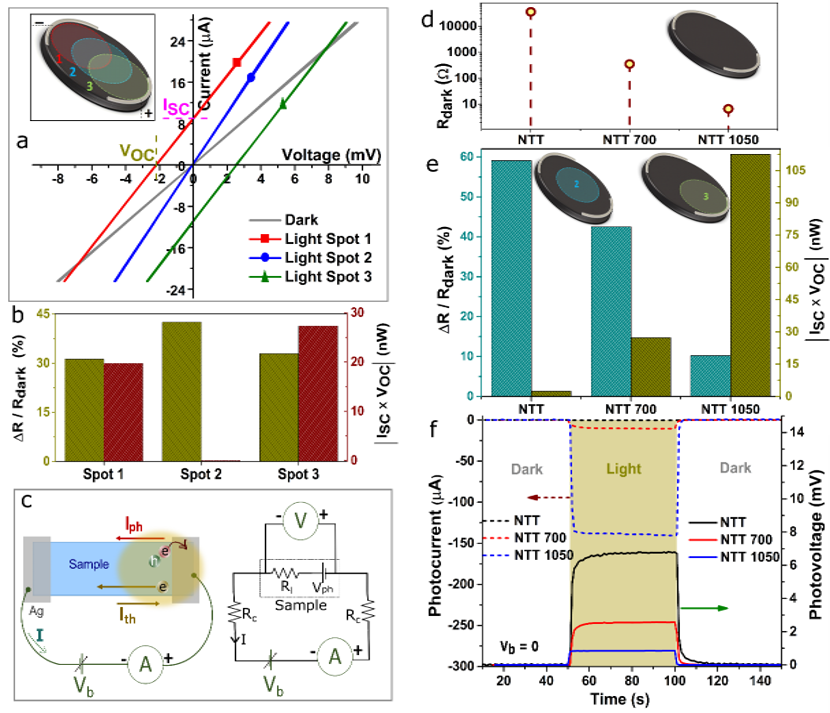}
		\end{center}
		\caption{Optoelectronic response of the systems. (a)$IV$ plots of sample NTT 700 in dark as well as with different illumination spots under white light illumination (power $\sim$0.8 W)[Inset; Schematic of the electrical setup for photo-response measurements]. (b) Photo-conductance(left) and Photovoltaic power(right) with corresponding illumination spots(for Sample NTT 700) and (c) proposes the mechanism of  observed photo-response. Comparison across samples are presented as (d) resistance in the dark, $R_{dark}$ (e)Photo-conductance(left) \& Photo-voltaic power(right) obtained from $IV$s and (f) shows zero bias (V$_b$=0) Photo-current(left) \& Photo-voltage(right) data with illumination at spot 3.}
		\label{fig:Elec}
	\end{figure}

The optoelectronic response of semiconductors originate from inter- and intra-band electronic excitations under photoexcitation that creates a non-equilibrium carrier distribution and the ensuing electron-phonon interactions, that are manifest in  the photoconducting, photovoltaic and finally thermophotoelectric signatures \cite{sett2022engineering}. In order to comprehend the optoelectronics of these novel reduced eutectic systems the photoresponse of NTT, NTT700 and NTT1050 were investigated.  All electrical measurements were conducted with extended metal (Ag paint) contacts along opposite edges of the circular samples as shown in the inset of fig. \ref{fig:Elec}(a). 
	 
	Figure \ref{fig:Elec}(a) shows the current-voltage, $IV$, characteristics for NTT700 in the dark and under white light illumination (power $\sim$ 0.8 W) focussed into a 1 mm diameter spot, at 3 different locations on the sample (inset fig. \ref{fig:Elec}(a)). The  $IV$s show linear, non-rectifying behavior with a decrease in resistance ($\Delta R/R_{d}\sim$ 40\%) upon illumination. Where $\Delta R$ is the difference between the dark resistance ($R_d$) and resistance with light ($R_l$). Asymmetric illumination, with light focussed around the contacts (spots 1 and 3) maximizes the open circuit photovoltage $V_{oc}$ and short circuit current $I_{sc}$ with maximum power ($P_{max}=\frac{1}{4}$($I_{sc}\times V_{oc}$)) as shown in fig. \ref{fig:Elec}(b).
	Samples NTT and NTT1050 show similar linear $IV$ characteristics as shown in figs. S11(a-b). 
	Fig. \ref{fig:Elec}(e) plots the $\Delta R/R_{dark}$ (left axis) for all samples for symmetric illumination at spot 2  along with $I_{sc}\times V_{oc}$ (right axis) under asymmetric illumination at spot 3. The temporal variation of $I_{sc}$ and  $V_{oc}$ with illumination at spot 3 is shown in fig. \ref{fig:Elec}(f). Together, fig. \ref{fig:Elec}(e) and fig. \ref{fig:Elec}(f) show that the least conducting NTT sample shows  the highest photoconductivity and $V_{oc}$ in contrast to the most conducting NTT1050 that exhibits the lowest photoconductivity and the highest $I_{sc}$. The significant absorption over the visible not only contributes to $e$-$h$ generation but also induces nontrivial photo-thermal effects. Here, the overall energy scale of the samples is set by (i) the electron affinity of TiO$_{2}$,  $\chi_e^{TiO_2}\sim$ 4.2 eV and  band gap $E_g\sim$ 3.5 eV \cite{yip2012recent} with the Magn$\acute{\mathrm{e}}$li phases (m\ch{TiO2}) having still lower $\chi_e$ and $E_g$ \cite{naldoni2012effect}, (ii) $\chi_e^{\ch{NiTiO3}}\sim$ 3.5 eV and $E_g\sim$ 2 eV \citealp*{yi2021heterostructure} and (iii) the work functions of Ni and Ag \cite{michaelson1977work}, $\phi_e^{Ni} \sim$ 5.01 eV and  $\phi_e^{Ag} \sim$ 4.4 eV, as shown in fig. S21. 
	In the phase segregated NTT sample,	the \ch{TiO2} phase is relatively more $n$-type and electron rich than the \ch{NiTiO3} phase, creating a system of distributed $pn$-like junctions at the phase boundaries. While band bending at the interfaces facilitate separation of photo-generated carriers and  the range of band gaps in the phases broaden the spectral absorption and contribute towards the $\sim 60\%$ photoconductance that is higher than the other samples (NTT700 and NTT1050). 
	In contrast, NTT1050 is a distributed Schottky junction system, formed between Ni NPs and m\ch{TiO2}. Ni on  $n$-type \ch{TiO2} forms a typical Schottky barrier, as shown in fig. S22 with $\phi_B\sim|\phi_e^{Ni}-\chi_e^{TiO_2}|\sim$ 1.0 eV \cite{michalas2018electrical}. The band alignment shown in fig. S21 indicates that Ag will form a low barrier height junction directly with m\ch{TiO2} and ohmic contact with Ni  which are reflected in the linear $IV$s in fig. S11b.     
	Further, NTT1050 shows activated transport ($R(T)\propto\exp (E_A/kT)$) with the activation energy $E_A\sim$ 10 meV  and a resistance of a few ohms at 300 K (see section S4.3.2 and fig. S19).
	To comprehend charge transport across the Ni-\ch{mTiO2} Schottky junction, especially under illumination, finite element method modelling was employed (see section S5.2 in SI). 
	The conducting nature of m\ch{TiO2} is mimicked via doping the \ch{TiO2} system (dopant density $\sim 10^{20}/cc$) in contact with the Ni globules.  In the dark, equilibrium is established as shown in fig. S23a with  upward band bending ($\sim 0.7 eV$) in \ch{TiO2} and a $\phi_B \sim 0.8eV$ that depletes \ch{TiO2} of electrons at the junction. Under super bandgap illumination the electron density in \ch{mTiO2} increases with increasing intensity (fig. S23d) causing a reduction in the sample resistance (fig. \ref{fig:Elec}e). 

	Central to the photovoltaic effect is the asymmetric illumination of the devices. Notably, NTT shows a higher photovoltage ($V_{oc}$) compared to NTT1050 but the lowest photocurrent (fig. \ref{fig:Elec}f) ensuring a higher power generation capability for NTT1050 (fig. \ref{fig:Elec}e).
	 The response though is not attributable to the Ag contacts  but is a ``bulk" effect arising from the photoresponse of \ch{mTiO2} amplified by the distributed $pn$ junctions in NTT and Schottky junctions in Ni-m\ch{TiO2}. Fig. S12a shows the $IV$ characteristics recorded with an illumination spot size $\sim$ 0.5 mm scanned across the electrodes on NTT1050. While $R_l$ changes little with the position of the spot, both $V_{oc}$ and $I_{sc}$ show systematic variation and are least for the spot equidistant from either contact, as plotted in fig. S12b. Asymmetric illumination has been shown to generate photovoltage in symmetric devices in which the local temperature and the nature of the junction at the contacts play crucial roles in determining response \cite{buscema2013large,omari2011photothermovoltaic,li2023reconfigurable,yang2019bolometric}.  
	Fig. \ref{fig:Elec}c shows the schematic of the symmetric Ag-NTT1050-Ag device illuminated at the right electrode (spot 3). Under zero external bias ($V_b$) the net current $I=I_{sc}$ is negative and flows from the left to right contact in the external circuit and in reverse through the device with electron flow in the opposite direction. $I$ changes direction  if the illumination spot shifts to the left electrode. 
	Fig. \ref{fig:Elec}c also shows the equivalent circuit where the sample with illumination is denoted by the resistor $R_l$ and a voltage source $V_{ph}$ that drives a photocurrent $I_{ph}$.
	The asymmetric illumination would also locally heat the sample creating a $T$ gradient inducing a thermal current ($I_{th}$) from left - right within the sample i.e. electrons flowing from high (right) to low (left) $T$, opposing $I_{ph}$.  Thus the net current ($I$) is the sum of currents induced by $V_b$, $I_{ph}$ and $I_{th}$, which under illumination at the right electrode is given by eqn.  \ref{eqn:cktcurr}, following the configuration shown in fig. \ref{fig:Elec}c. 
	\begin{equation}
	I=(V_{ph}+V_b)/(R_l+2R_c)-I_{th}
	\label{eqn:cktcurr}
	\end{equation} 
	where $R_c$ is the contact resistance with Ag, and $I_{sc}=I_{ph}-I_{th}$ with $I_{ph}=V_{ph}/(R_l+2R_c)$. Direction of the $I_{sc}$ shows that $I_{th}<I_{ph}$ though not negligible and $I_{ph}$ dominates the net current in the circuit. The role of the Ni NPs and the distributed Schottky junctions present in  NTT1050 but absent in NTT is elucidated by the FEM simulations of the  Ni-\ch{mTiO2} junction system. Under illumination the photoexcited electrons in \ch{mTiO2}  repopulate the depleted conduction band states at the Ni interface increasing the interfacial $n_e$ by orders in magnitude. SI fig. S23d shows increase in $n_e$ of over four orders in magnitude with increase in incident radiation intensity. The accompanying decrease in the built-in potential shown in SI fig. S23c and reduced band bending (SI fig. S23a) along with the weakening junction field increases electron transfer from \ch{mTiO2} to Ni, especially in a closed circuit configuration. Thus, the Ni NPs  dispersed across the m\ch{TiO2} network, under optical excitation aid photogenerated electron extraction from the semiconductor and facilitate transport across the system via the Ag contacts.

\begin{figure}
		\begin{center}
			\includegraphics[width=0.45\textwidth]{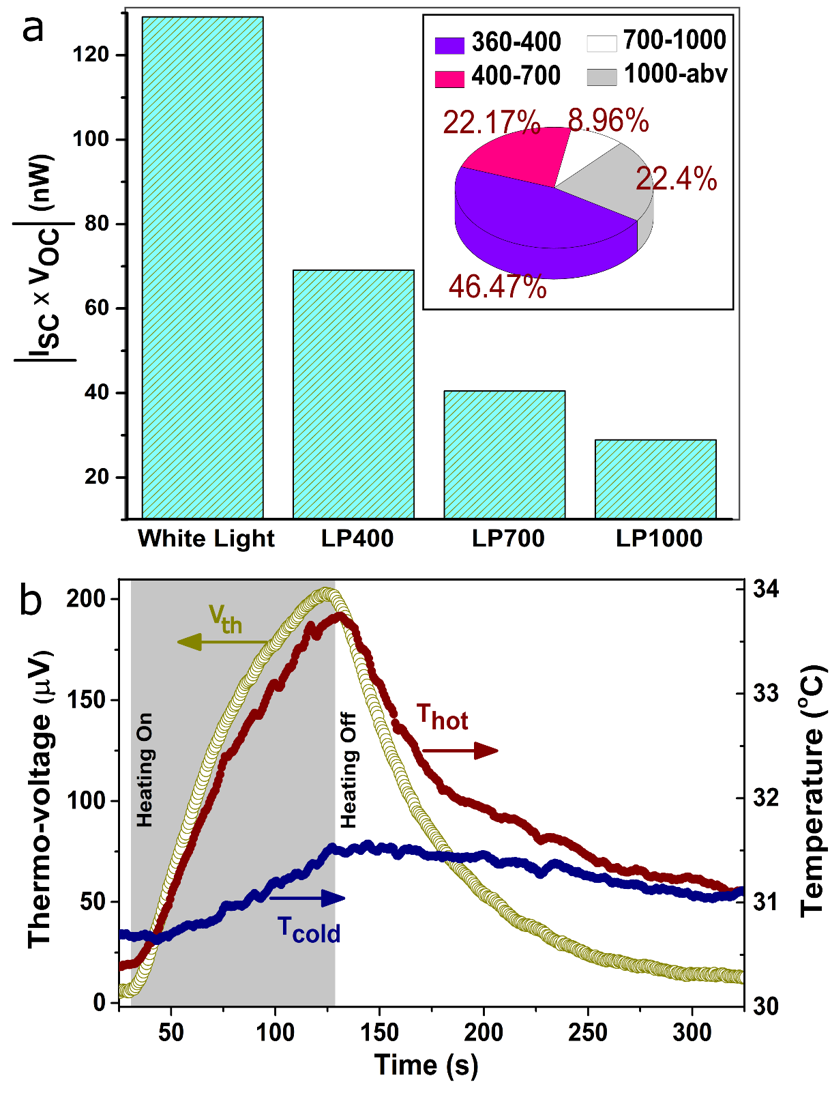}
		\end{center}
		\caption{Photo-thermo-voltaic characteristics of NTT1050. (a)Photovoltaic performance with illumination with various long pass (LP) optical filters (LP400, LP700, LP1000 filters with cut-on wavelength at 400, 700 and 1000 nm, respectively), inset shows percentage power generated in different spectral ranges. (b) thermo-voltage $vs$ time plot}
		\label{fig:Comp}
	\end{figure}

The photoresponse of NTT1050 ensues from its unique architecture and varied stoichiometry.  
The \ch{mTiO2} harbours a rainbow of band-gaps that widen the spectral absorption range and the distributed Ni junctions facilitate electron transport, especially under illumination make it a promising photovoltaic material atop a conducting backbone\cite{he2012enhanced}.  To investigate  the broadband nature of the photoresponse, fig. \ref{fig:Comp}(a) plots $|I_{sc}\times V_{oc}|$ for illumination with various long pass (LP) optical filters, with the corresponding $IV$s shown in fig. S13. Inset of fig. \ref{fig:Comp}(a) shows the fraction of the total photovoltaic power attributed to different spectral windows. More than 50\% of the photovoltaic  power is generated from irradiation above 400 nm i.e. energies lower than the bandgap of \ch{TiO2}, which accentuates the role of the Magneli and sub-oxide phases that increase absorption and contribute to overall increase in $n_e$ and thus the photoelectric response i.e. $I_{ph}$ and $V_{ph}$. Notably, $\sim$ 22\% power arises from radiation above $1000$ nm indicating significant absorption and carrier generation in that regime. The remaining photoresponse arises from radiation below 400 nm, corresponding primarily to super-bandgap excitation of \ch{TiO2}.  The observations are corroborated by the spectrally resolved power ratio plot shown in fig. S14. 
	
	\subsection*{Thermal \textit{vs} Optical response}
	Further insights into the thermal current and emf may be obtained from a direct thermo-emf measurement. The highly porous architecture of NTT1050 with the m\ch{TiO2} backbone makes the system conducive for thermo power applications, though the high electrical conductivity remains a drawback. 
	A custom made thermo-emf measurement set-up, as shown in the schematic in fig. S16a was used to measure the thermo-emf across the device  (see SI section S4.2 for further details). Fig. \ref{fig:Comp}(b) shows the $T$ variation of the hot and cold heads as a function of time along with the variation of thermo-emf ($V_{th}$). 
	 The calculated Seebeck coefficient S $\sim$ 70 $\mu$V/K, at room temperature  (fig. S16b) is unusually high for a material with similar conductivity \cite{cougnon2019seebeck,mateeva1998correlation} and comparable to the values reported for m\ch{TiO2}\cite{ovsyannikov2012pressure}. 
	The potential for titanium sub-oxides and the Magn$\acute{\mathrm{e}}$li phases in thermoelectrics stem from the crystallographic shear plane that  is selectively a strong phonon scatterer but not of electrons\cite{harada2010thermoelectric,liu2017high}.
			\begin{figure}
		\includegraphics[width=0.9\textwidth]{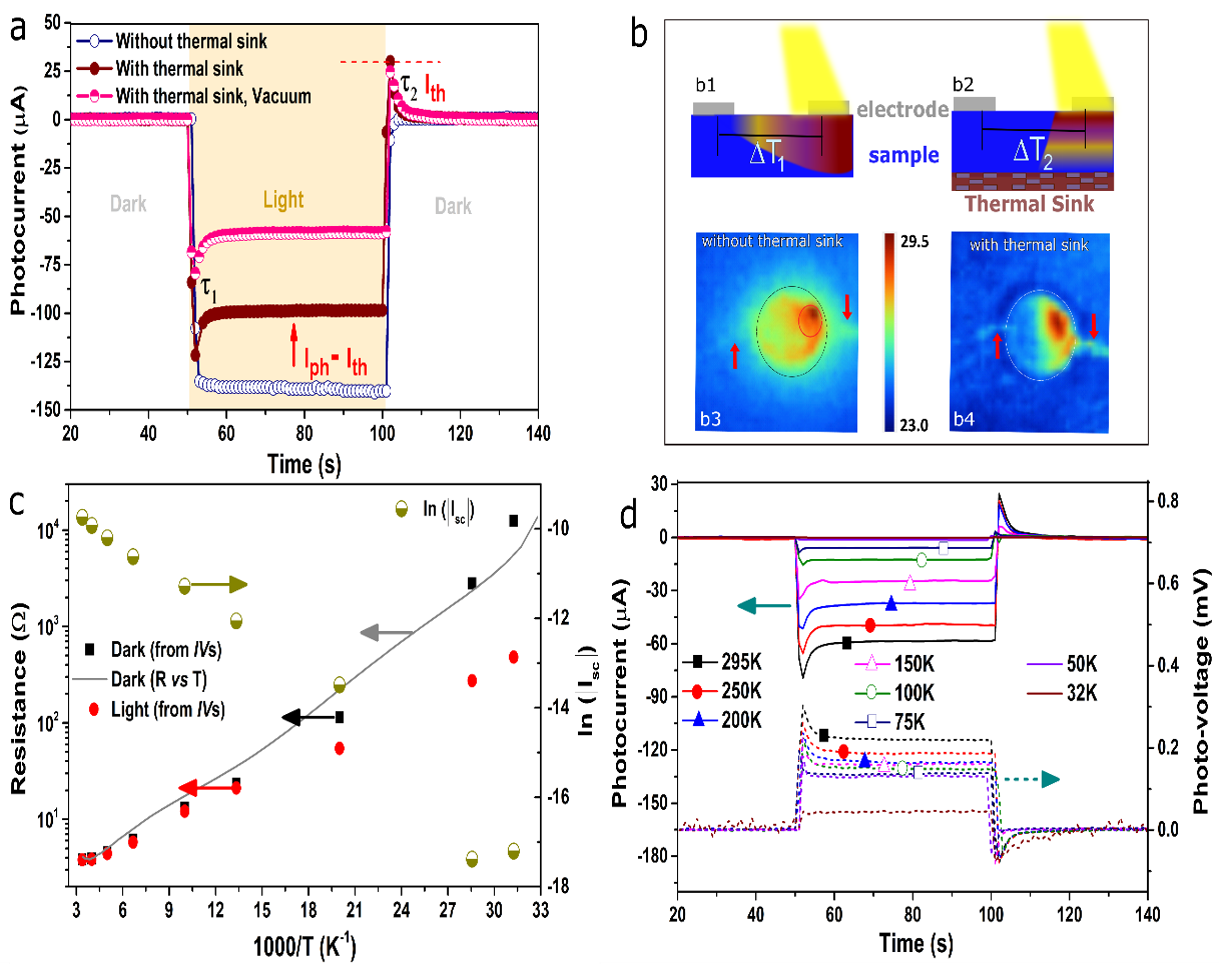}
		\caption{(a) Temporal response of $I_{sc}$, evidencing the effect of asymmetric illumination on and off, with the device without a heat sink, with a heat sink and under vacuum, (b) top: Schematic of device and induced temperature gradient without and with a heat sink, bottom:  Thermal maps showing temperature variation across sample under both condition  (black circle outlines the sample edge, red circle denotes illumination spot and red arrows denote electrical wires). (c) Variation of resistance(left) and photo-current(right) with temperature and (d) shows the temporal variation of zero bias photo-current and photo-voltage as a function of temperature.}
		\label{fig:LT}
	\end{figure}
	The $I_{th}$ term in eqn. \ref{eqn:cktcurr} and associated photo-thermal effects under photo-excitation assumes significance in view of the large value of S.  
	Comparison of the temporal variation of the thermo-emf $V_{th}$ and $V_{oc}$ under photo-excitation  plotted in figures S17(a) and (b) show that the two processes have very different time scales. The $V_{oc}$ dominated by the photo-voltaic response has a rise time constant, $\tau_{r1}\sim$ 0.6 s and a decay time constant $\tau_{d1}\sim$ 0.4 s. The $V_{th}$ kinetics yield $\tau_{r2}\sim$ 64 s and $\tau_{d2}\sim$ 49 s, two orders in magnitude slower than the photo-response.   Under asymmetric illumination, an estimate of the fractional contribution of $I_{th}$ to the total current is obtained if the experiment (fig. \ref{fig:Elec}(f)) is repeated with the device fixed to a copper heat sink.    
	Fig. \ref{fig:LT}(a) plots the  variation in $I_{sc}$ as illumination is turned on and off at spot 3 on the device without any heat sink, with a heat sink and with a heat sink in vacuum. 
 Without a heat sink  $|I_{sc}|$ increases monotonically 	upon illumination and saturates with a time constant $\tau_{r1}$ (fig. S17b). However, when placed on a heat sink $|I_{sc}|$ increases rapidly, reaches a maximum and then decreases at a slower rate, with a  $\tau_1\sim$ 2 s to saturate at a steady state value. Both the maximum and the saturation values of $|I_{sc}|$ under illumination with the heat sink are smaller than those recorded without the heat sink. 
	Once illumination is turned off, $I_{sc}$ rapidly decays and saturates at zero with the time constant $\tau_{d1}$ (fig. S17b) in the case without a heat sink. However, with a heat sink $I_{sc}$ decreases towards zero but does not saturate and overshoots to evidence a current in the opposite direction, reaches a maximum and then decays to zero with $\tau_2\sim$ 2 s. 
	Both effects are replicated with the device on heat sink in an evacuated environment (10$^{-3}$ mbar vacuum), though the  $|I_{sc}|$ saturation value upon illumination is further lower than the case with heat sink without vacuum. The  lower saturation $|I_{sc}|$ with the device on heat sinks (in ambient and in vacuum), and the overshoots above saturation are measures of $I_{th}$ that is induced due to photothermal heating and flows in the opposite direction to $I_{ph}$.
	Fig. \ref{fig:LT}(b1\&b2) shows schematics of the device illuminated at the right electrode (spot 3) without and with a heat sink and the ensuing temperature gradient across the sample. 
 For illumination at spot 3, the right electrode would have higher local temperature $T_h$ than the left electrode ($T_c$) creating a temperature difference $\Delta T=T_h-T_c$ across the sample. Figs. \ref{fig:LT}(b3\&b4) show the infrared thermal maps recorded for the system without and with heat sink, in the ambient. Both maps were recorded immediately after illumination for 60s and show that $\Delta T$ is higher in the case with heat sink than without the sink. Though the maximum $T_h\sim 29.5^oC$ reached is either case is comparable,  the thermal gradient ($dT/dx$) across the sample is half for the sample without the heat sink than with the heat sink, as shown in fig. S18c (SI). 
Consequently, the local free electron chemical potential ($\mu_e$) would be higher at the hotter end (right electrode) inducing an $I_{th}$ within the sample from left - right, that opposes the $I_{ph}$ flowing from right to left through the sample. Thus, the presence of the heat sink increases $dT/dx$ inducing a higher $I_{th}$. 
Since $I_{ph}$ increases faster than $I_{th}$, with the device on heat sink the initial rise of $I_{sc}$ (with $I_{ph}$) is arrested by the slower rising $I_{th}$ limiting $I_{sc}$ to a maximum which then decays to saturate. 
Once the illumination is turned off the $I_{ph}$ decays much faster than $I_{th}$ thus the net current transitorily becomes opposite in direction before decaying to zero. Heat loss to the environment is further reduced in vacuum,  which increases $dT/dx$ across the sample thereby increasing $I_{th}$ and lowering  the saturation current value of $|I_{sc}|$.

\subsection*{Temperature dependent measurements}
	The thermally activated nature of electrical transport evidenced in NTT1050 with $E_A \sim$ 10 meV offer further insights to its photoresponse.  Fig. \ref{fig:LT}c shows variation in device resistance with $T$ in the dark evidencing more than 3 orders in magnitude increase in resistance from 300 - 30 K, along with a scatter plots of resistance calculated from $IV$s in the dark and under asymmetric white light illumination. 
	While the photoconductance of the sample increases with decreasing $T$  (see SI fig. S20), the saturation values of $|V_{oc}|$ and $|I_{sc}|$, both decrease with decreasing temperature thus compromising the power generation capability.
	Fig. \ref{fig:LT}d shows  with temporal variation of $I_{sc}$ and $V_{oc}$ recorded with asymmetric illumination on the sample on a heat sink in vacuum, at various temperatures.  Further, the $T$ dependence of $|I_{sc}|$ shown in fig. \ref{fig:LT}c attests to a functional dependence $\ln(I_{sc})\propto -1/T$ that is characteristic of the photovoltaic response of diodes \cite{dalapati2018effect}. 
Lowering temperature would decrease the free carrier density in \ch{mTiO2}, finally leading to carrier freeze out below 100 K. The concomitant increase the $\phi_B$ and resistance of the Ni-m\ch{TiO2} junctions indicate electrical isolation of the \ch{mTiO2} backbone from the Ni NPs at lower $T$. This progressively reduces the effect of the distributed Schottky junctions in inducing the photovoltaic properties of NTT1050, as evidenced through the $T$ dependence of $I_{sc}$. The temperature dependence further attests to the central role of the distributed junctions in the overall photovoltaic response of \ch{mTiO2}. 

	
	\section{Conclusions}
	In summary, we have investigated the photoresponse of a Ni - 
	m\ch{TiO2} hybrid with a unique morphology where Ni nanoparticles decorate a conducting dendritic 
	m\ch{TiO2} network. The optoelectronic response evidence photovoltaic, thermoelectric and photothermal effects in two-terminal devices that are surprising given the conducting nature of the system. Crucially, photovoltage in the hybrid does not arise from contact junction effects, which are ohmic in nature but is a "bulk" effect arising from distributed Ni - Magn$\acute{\mathrm{e}}$li \ch{TiO2} Schottky junctions that ensure carrier separation and transport, needed for photovoltaic power generation.  Asymmetric illumination of the highly porous, two terminal device may be used to build self powered sensors that are easily upscaled due to the simple architecture. 
	
	\section*{Methods}
	\subsection*{Synthesis and processing}
	\ch{NiTiO3 -TiO2 } rods were grown by the micro-pulling-down method ($\mu$-PD) with high-purity \ch{TiO2} and NiO powders as the starting materials in a nitrogen atmosphere (see S1 of SI for further details). After the growth, the composite rods were cut into small discs ($\sim$600 $\mu$m in thickness) perpendicular to the growth direction (Fig S1). Then the samples were polished, and finally, some of the samples were annealed in hydrogen (\ch{H2}) atmosphere (100\% flow) at high temperatures (700, 1050$^{\circ}$C).
	\subsection{Characterization}
	X-ray diffraction (XRD) data of powdered samples were collected using a powder X-ray diffractometer (Empyrean, PANalytical, Rigaku SmartLab(R) 3kW) with Cu K$\alpha$ radiation of 1.5418 \AA, at an operating voltage of 45 kV and analyzed using the ICDD PDF4+2018 database. Morphological and elemental analysis were performed using a Nova Nano SEM 450, field-emission scanning electron microscope (SEM) coupled with Apollo X (EDAX Inc.) energy dispersive X-ray analysis (EDS) system. X-ray photo-electron spectroscopy (XPS) was measured in a Scienta Omicron XPS to understand the oxidation states and thereby the chemical state of our sample. The sample surface was cleaned by an in-situ Argon etching prior to obtaining XPS data to avoid surface contaminants. Raman spectra were obtained by using a Horiba confocal micro Raman Xploraplus spectrometer using 532 nm excitation. A Bruker Multimode 8 Atomic Force Microscope (AFM) was used to characterize their surface morphology and the height profile.
	\subsection{Optoelectronic measurements}
	Pseudo four-probe contacts, extended metal (Ag paint, Ted Pella, Inc.) contacts along opposite edges of the sample with an active device area of 0.21 $\pm$ 0.03 cm$^2$ were made for conducting the electrical (resistance and current- voltage [$IV$] characteristics) and photo-response [$PR$] measurements. All electrical measurements were performed using a Keithley 2400 sourcemeter. The $IV$ characteristics were recorded by sourcing current and measuring voltage across the devices. Photoresponse ($PR$) measurements were carried out by illuminating samples (focused to $\sim$1 mm diameter) with a Stabilized Quartz-Tungsten Lamp (Thorlabs SLS301), with band-pass, neutral density (ND), $\lambda$ and long pass (LP) filters added for varying intensity and to record spectral dependence of PR. The optical power of illuminations are measured using a Thorlabs powermeter(PM100D) and photodiode power sensors (S122C \& S120VC).
	\subsection{Temperature dependent measurements}
	The thermo-voltaic measurements were done using Keysight DAQ6510 in a custom-made probe station. The samples are fixed using Ag paint to two copper blocks kept side by side, where one of the copper blocks is equipped with a cartridge heater. Thermocouples and electrical probes are attached to the sample sides to monitor the change in voltage with temperature ($\Delta$V \& $\Delta$T). See S4.2 of SI for further details.\\
	Low-temperature opto - electrical measurements were performed with the sample mounted on a closed cycle cryostat (ARS Inc.) with GE varnish, connected to a Lake Shore 336 temperature controller.
	\subsection{Finite Element Modelling}
	Finite element modelling of a Schottky junction between Ni and \ch{TiO2} was carried out using COMSOL 6.0 in the semiconductor module to observe and analyse the band characteristics and electron distribution in such a system.  See S5.2 of SI for further details.
	
	\begin{acknowledgement}	
    The authors thank Prof. Hema Somanathan (IISER TVM) for access to the thermal imaging facility and Dr. Soumya Biswas (IISER TVM) for help in conducting the thermoelectric measurements. The authors acknowledge financial support from DST, Govt. of India (DST/INT/POL/P-44/2020) and NAWA Bilateral exchange of scientists (PPN/BIN/2019/1 /00111). Authors  acknowledge the support from Office of the PSA, Govt of India for COMSOL access through I-STEM program.  HG acknowledges PhD fellowship from DST INSPIRE, Govt. of India and SC, AK acknowledges IISER TVM for the PhD fellowship. KB, KK and DAP thank the financial support from ENSEMBLE$^3$ Project (MAB/2020/14), which is carried out within the International Research Agendas Programme (IRAP) of the Foundation for Polish Science co-financed by the European Union under the European Regional Development Fund and Teaming for Excellence Horizon 2020 programme of the European Commission (GA No. 857543).
	\end{acknowledgement}
	
\section{Author Contributions}
HG, JM conceptualized the investigation, KB, KK, DAP synthesized the samples. HG, AK, VBK conducted various experiments and performed the analysis and SC conducted the simulations. All authors contributed to writing the manuscript.

\section{Conflicts of interest}
The authors declare no competing financial interest.

\begin{suppinfo}
Synthesis,experimental procedures and characterization data for all new compounds. Photoresponse measurement details; illumination spectra and power, and spacial and spectral resolved photoresponse. Thermo-emf measurement set-up and details. Low temperature electrical measurements data. Energy levels of components and band alignments. Finite Element Modeling details and results.		
\end{suppinfo}

\clearpage
\appendix
\renewcommand{\theequation}{S\arabic{equation}}
\renewcommand{\thetable}{S\arabic{table}}
\renewcommand{\thefigure}{S\arabic{figure}}
\renewcommand{\thesection}{S\arabic{section}}
\setcounter{equation}{0}
\setcounter{table}{0}
\setcounter{figure}{0}
\setcounter{section}{0}

\section*{Supporting Information (SI)}
\addcontentsline{toc}{section}{Supporting Information (SI)}

	
\section{Synthesis}
\begin{figure}[H]
	\begin{center}
		\includegraphics[width=0.67\textwidth]{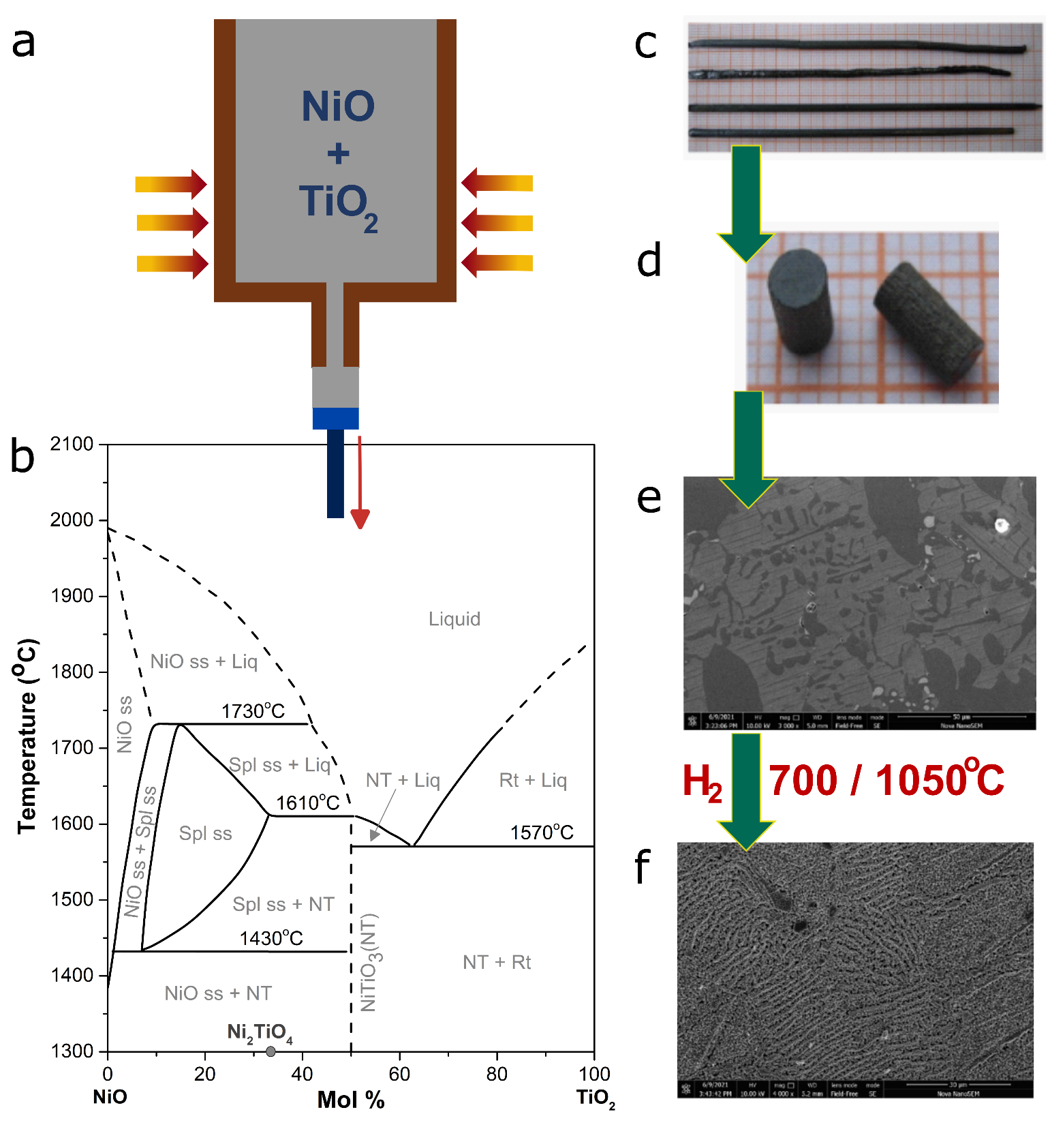}
		\caption{Scheme of Synthesis. (a) $\mu$-PD method (b) Phase-diagram of NiO-\ch{TiO2}\cite{muan1992equilibrium} (c) optical image of synthesized eutectic rod (d)perpendicularly cut piece. SEM images of (e) polished eutectic surface and (f) $1050^oC$ H$_2$ annealed sample.}
		\label{fig:S1}
	\end{center}
\end{figure}

\ch{NiTiO3}-\ch{TiO2} rods were grown by the micro-pulling-down method (µ-PD)\cite{fukuda2004fiber,yoon1994crystal,kolodziejak2016synthesis} in a nitrogen atmosphere following the phase diagram as in fig. \ref{fig:S1}(b)\cite{muan1992equilibrium}. High-purity \ch{TiO2} (99.995\%) and NiO (99.999\%) powders (Alfa Aesar) were used as the starting materials. Materials were mixed (with an over-eutectic composition of 69 mol\% \ch{TiO2} and 31 mol\% NiO) with pure 2-propanol in an alumina mortar. Over-eutectic composition was chosen to obtain more \ch{TiO2} in the system. The obtained slurry was then annealed at 90$^oC$ for 2 h to remove potential volatile impurities. Then the obtained mixture was put in a Nabertherm N20/H oven at 1100$^oC$ for 10 h in $N_2$ or $Ar/CO_2$ atmosphere. The final raw material was used for the growth of \ch{NiTiO3}-\ch{TiO2} rods in the $\mu$-PD Cyberstar apparatus. 
In the $\mu$-PD, the melts flows through the capillary and solidifies outside of the crucible die. The solidification starts after touching the melt at the die bottom with a seed (previously prepared \ch{NiTiO3-TiO2}). Next, the seed is retracted with constant pulling rate, which conditions uniformity of eutectic materials microstructure.  In the case of a round-shaped crucible die the material is obtained in the form of a rod. The applied pulling rate was 3 mm/min. After the growth, the composite rods were cut into small discs ($\sim$600 $\mu$m in thickness) perpendicular to the growth direction(figure\ref{fig:S1}(d)). 
Then the samples were polished and finally some of the samples were annealed in hydrogen ($H_2$) atmosphere(100\% flow) at high temperatures (700, 1050$^oC$). The annealing step initiates and causes the reduction of nickel titanate\cite{arvanitidis2000intrinsic, morales2005hydrogen} to nickel and titanium dioxide (Eq\ref{eqn:synth}).
\begin{equation}\label{eqn:synth}
	NiTiO_3 + H_2 \rightarrow Ni + TiO_2 +H_2O \\   (610^oC \le T \le 862^oC)
\end{equation}
Thus the nickel diffuses out and forms the NPs on TiO$_{2-\Delta}$ interconnected fibrous network structure (fig. \ref{fig:S1}(f)).

Hereafter samples \ch{NiTiO3}-\ch{TiO2} are labelled NTT and the \ch{H2} annealed samples are named NTT700 \& NTT1050, according to the annealing temperature.

\section{Characterization}
X-ray diffraction (XRD) data of powdered samples were collected using a powder X-ray diffractometer (Empyrean, PANalytical, Rigaku SmartLab(R) 3kW) with Cu K$\alpha$ radiation of 1.5418 \AA, at an operating voltage of 45 kV and analyzed using the ICDD PDF4+2018 database. Morphological and elemental analysis were performed using a Nova Nano SEM 450,
field-emission scanning electron microscope (SEM) coupled with Apollo X (EDAX Inc.) energy dispersive X-ray analysis (EDS) system. X-ray photo-electron spectroscopy (XPS) was measured in a Scienta Omicron XPS to understand the oxidation states and thereby the chemical state of our sample. The sample surface was cleaned by an in-situ Argon etching prior to obtaining XPS data to avoid surface contaminants. Raman spectra were obtained by using a Horiba confocal micro Raman Xploraplus spectrometer using 532 nm excitation. A Bruker Multimode 8 Atomic Force Microscope (AFM) was used to characterize their surface morphology and the height profile.

\subsection{Identification of phases in NTT}

\begin{figure}[H]
	\begin{center}
		\includegraphics[width=0.77\textwidth]{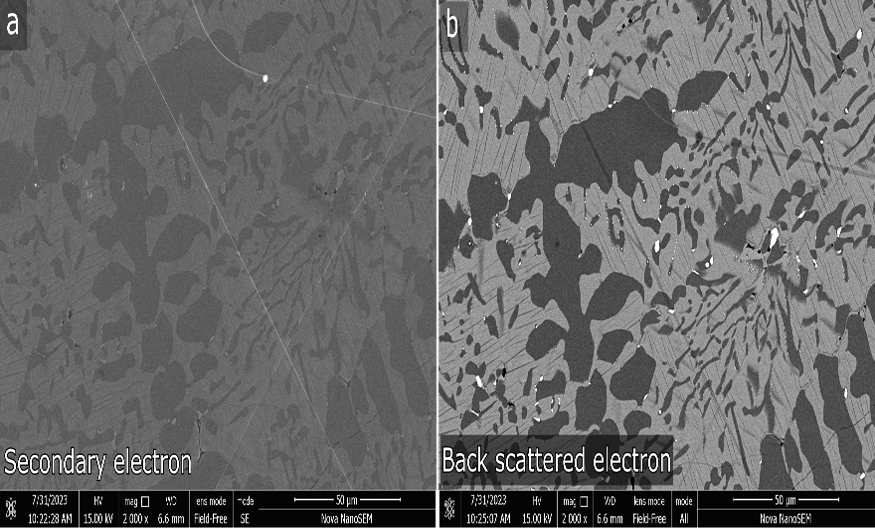}
		\caption{Comparison between (a) secondary electron and (b) back scattered electron SEM images of NTT.}
		\label{fig:S2a}
	\end{center}
\end{figure}
\begin{figure}[H]
	\begin{center}
		\includegraphics[width=0.9\textwidth]{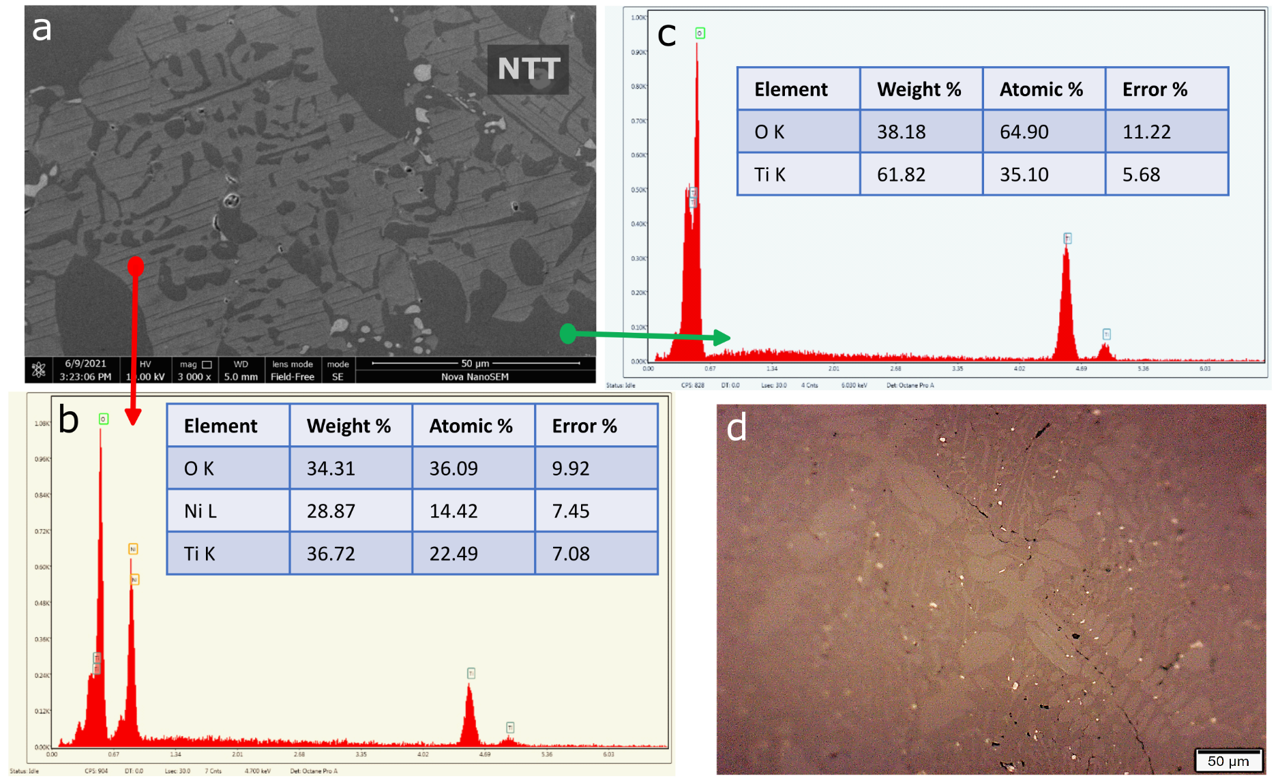}
		\caption{Elemental analysis in NTT. (a)SEM image of the polished NTT over-eutectic sample containing different phases(dark \& bright) with (b \& c) corresponding EDS data. The dark region consists of Oxygen deficient \ch{TiO2} and the bright region is of \ch{NiTiO3} phase. (d) Optical microscope image of NTT.}
		\label{fig:S2}
	\end{center}
\end{figure}

\begin{figure}[H]
	\begin{center}
		\includegraphics[width=0.6\textwidth]{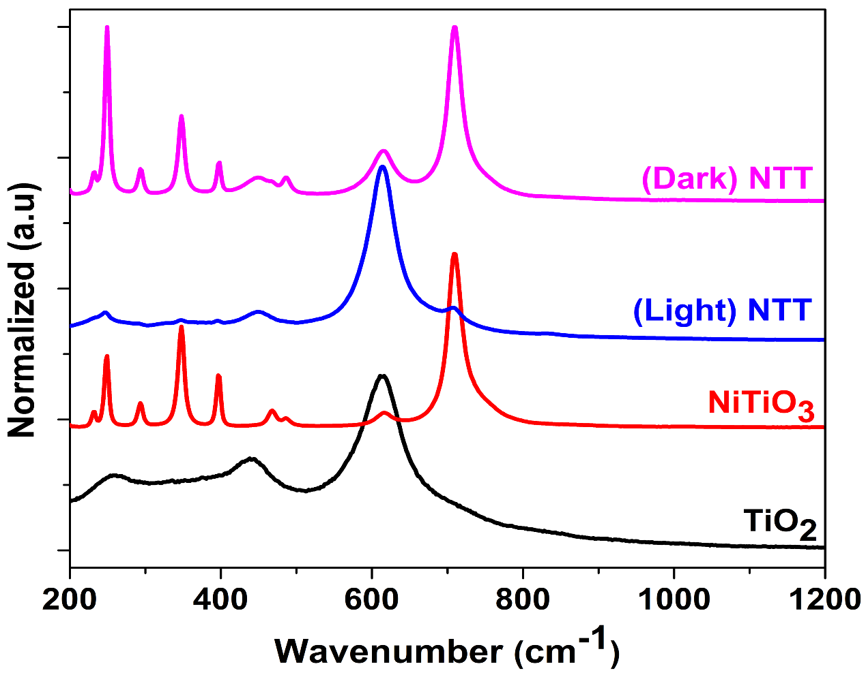}
		\caption{The Raman data of samples; \ch{TiO2}, \ch{NiTiO3} and dark \& bright portions of NTT.}
		\label{fig:S2b}
	\end{center}
\end{figure}

\subsection{Characterization of NTT1050}

\begin{figure}[H]
	\begin{center}
		\includegraphics[width=0.75\textwidth]{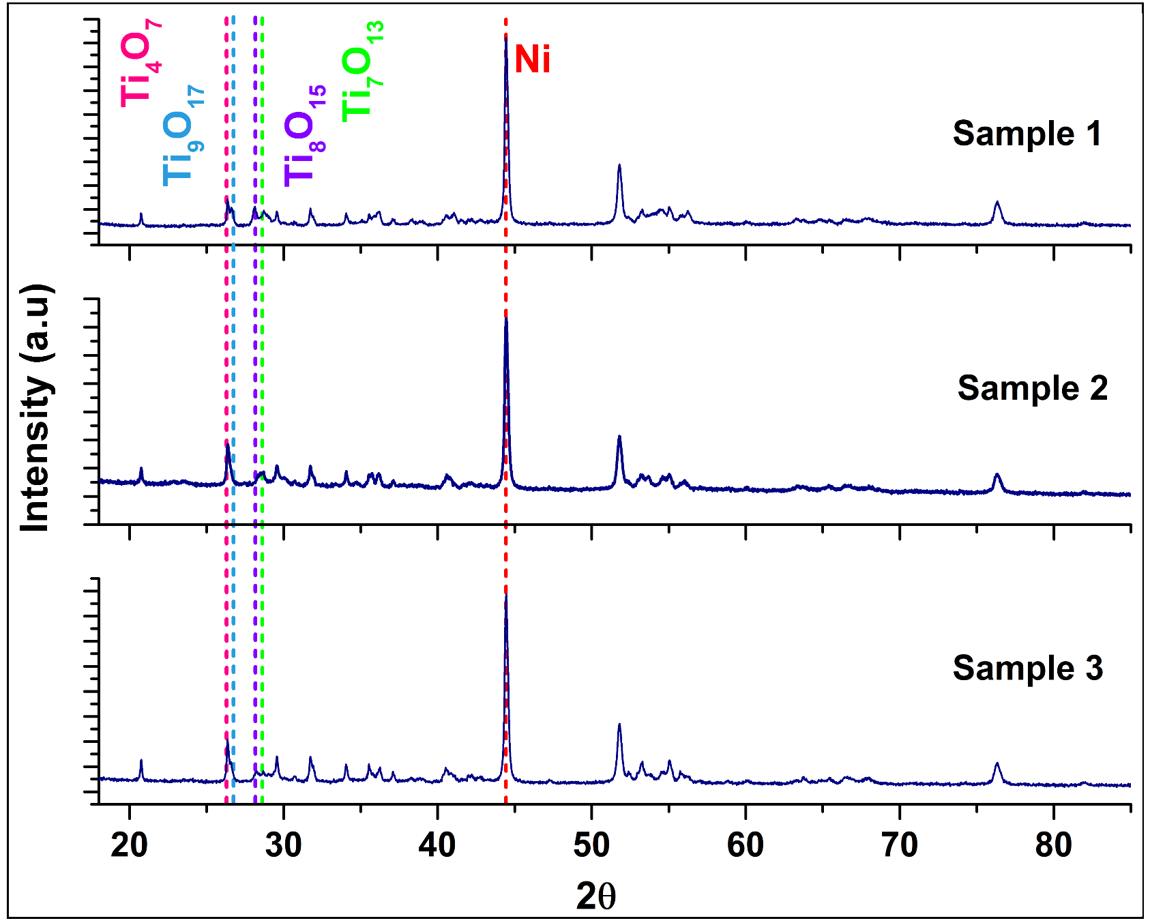}
		\caption{Powder XRD data of NTT1050.  Samples 1 to 3 shows different set of samples indicating the presence of various magneli phases in the samples.}
		\label{fig:S3}
	\end{center}
\end{figure}
\begin{figure}[H]
	\begin{center}
		\includegraphics[width=0.95\textwidth]{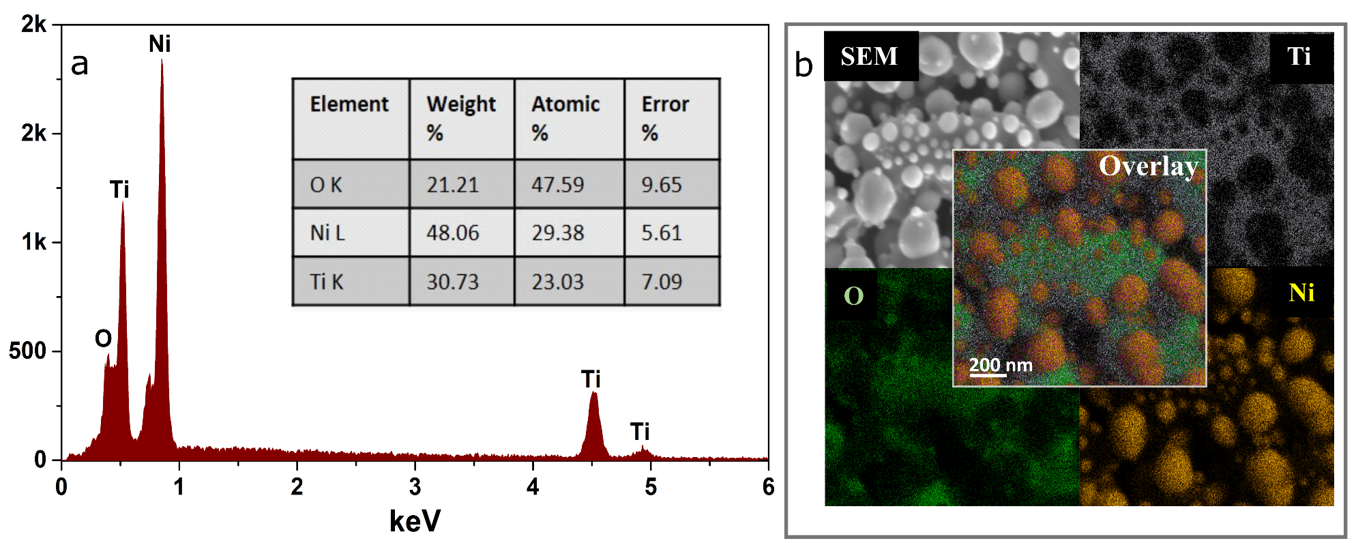}
		\caption{Energy dispersive spectroscopy of NTT1050 (elemental percentages in the inset), and (b) elemental mapping.}
		\label{fig:S3a}
	\end{center}
\end{figure}

\begin{figure}[H]
	\begin{center}
		\includegraphics[width=0.85\textwidth]{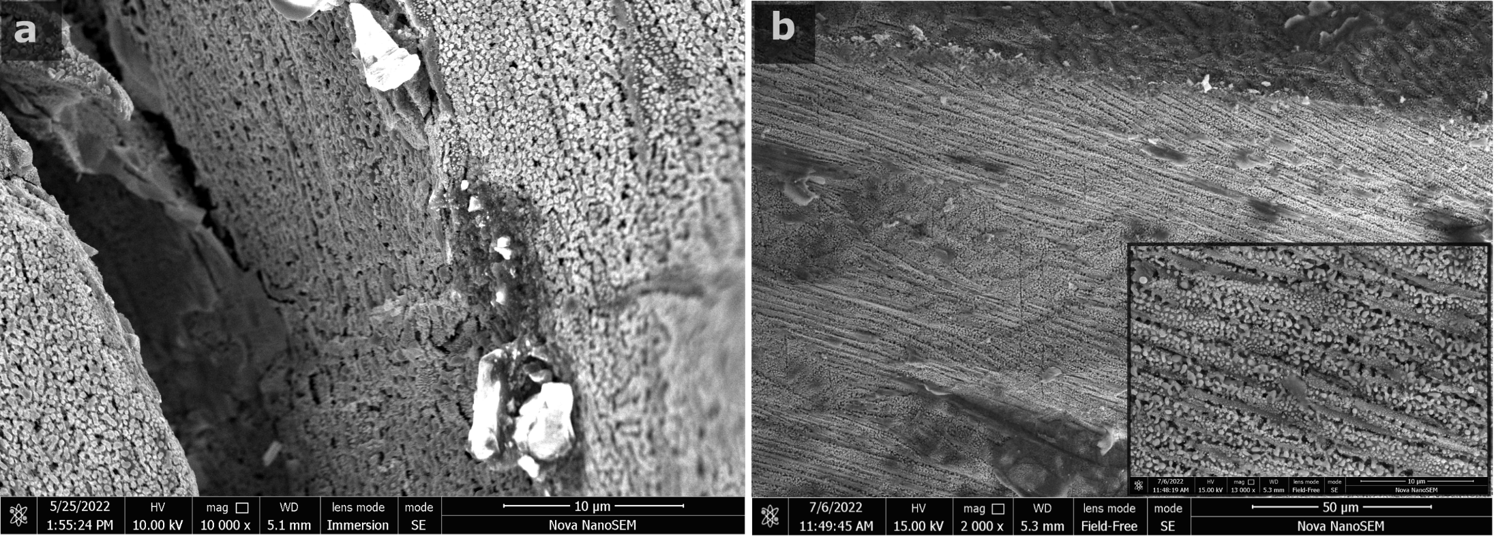}
		\caption{SEM images of (a)a crack on NTT 1050 and (b)a broken side of NTT1050 showing the continuity of morphology along the depth of the sample.}
		\label{fig:S3b}
	\end{center}
\end{figure}

\begin{figure}[H]
\begin{center}
	\includegraphics[width=0.95\textwidth]{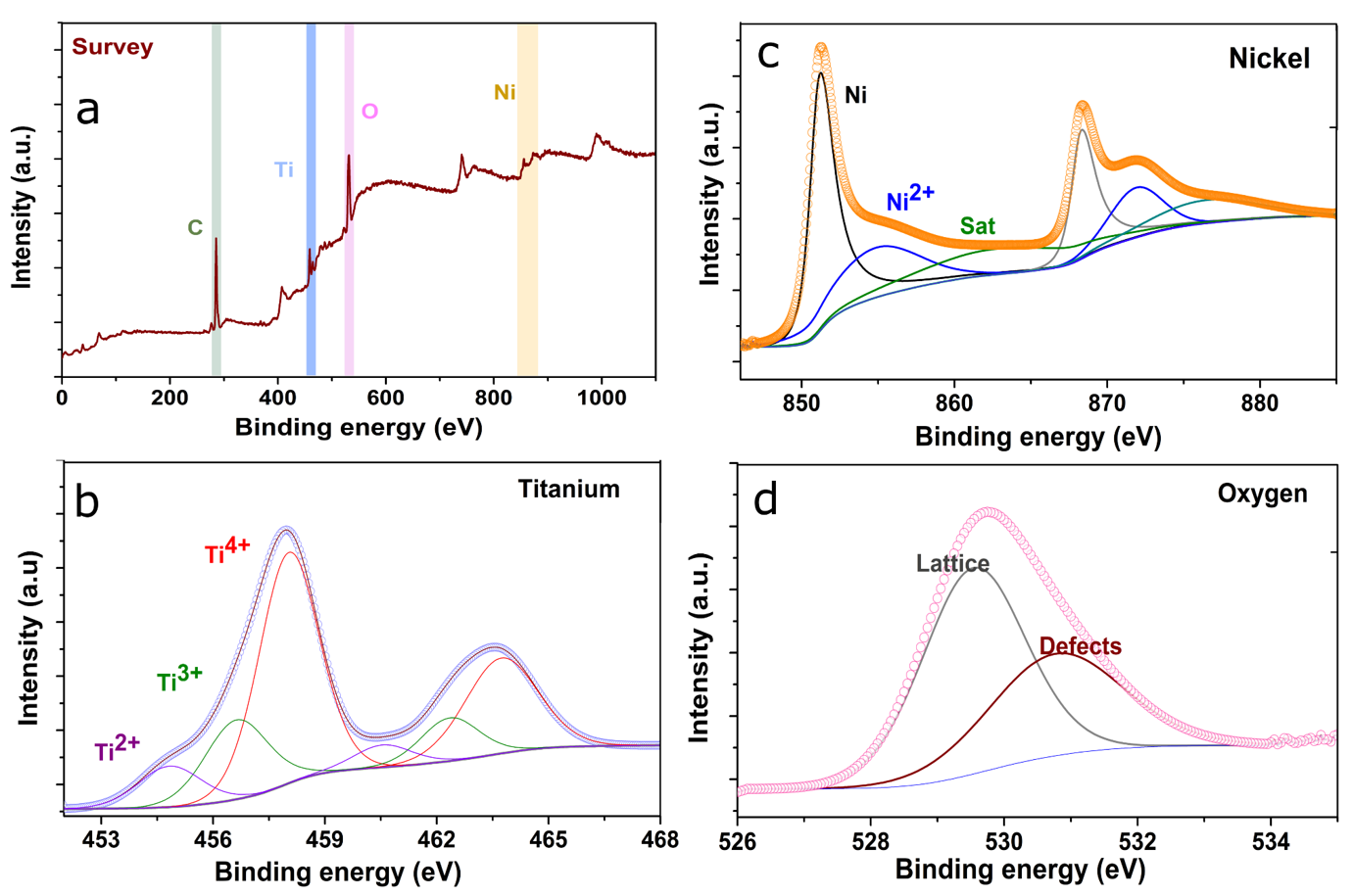}
	\end{center}
	\caption{XPS data of NTT1050 showing (a)survey scan and high-resolution (HR) spectra of (b)Ti 2p (c)Ni 2p \& (d)O 1s. The HR spectra (c and b) are deconvoluted to get peaks and are assigned with corresponding oxidation states (Sat- satellite peak). HR spectra of O1s (d) is deconvoluted and arranged such that the lattice (of different phases of Titanium Oxide) related peaks are enveloped under “Lattice” and “Defects” peak is attributed to various defects.}
	\label{fig:S3c}
	\end{figure}

\subsection{Optical properties}
The diffuse reflectance of the samples are obtained with an integrating sphere using Perkin Elmer Lambda 900 spectrophotometer. The samples were mounted on a spectralon disk that has the highest diffuse reflectance over the UV to IR spectrum.
\begin{figure}[H]
	\begin{center}
		\includegraphics[width=0.65\textwidth]{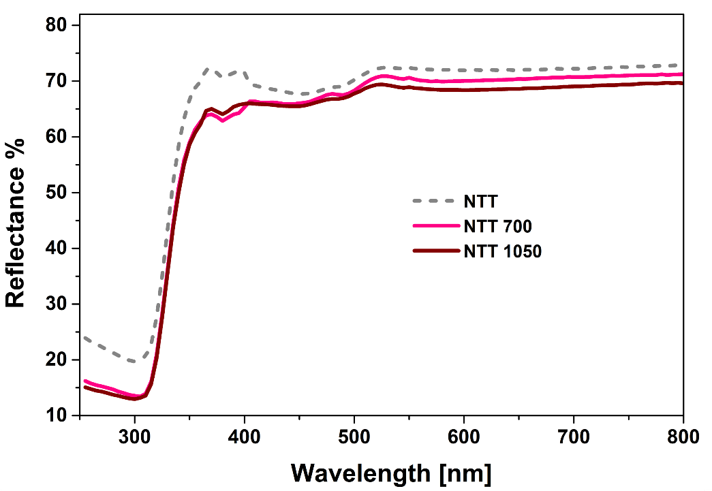}
		\caption{Diffuse reflectance data of samples.}
		\label{fig:S4}
	\end{center}
\end{figure}

\section{White light Lamp}
\begin{figure}[H]
	\begin{center}
		\includegraphics[width=0.95\textwidth]{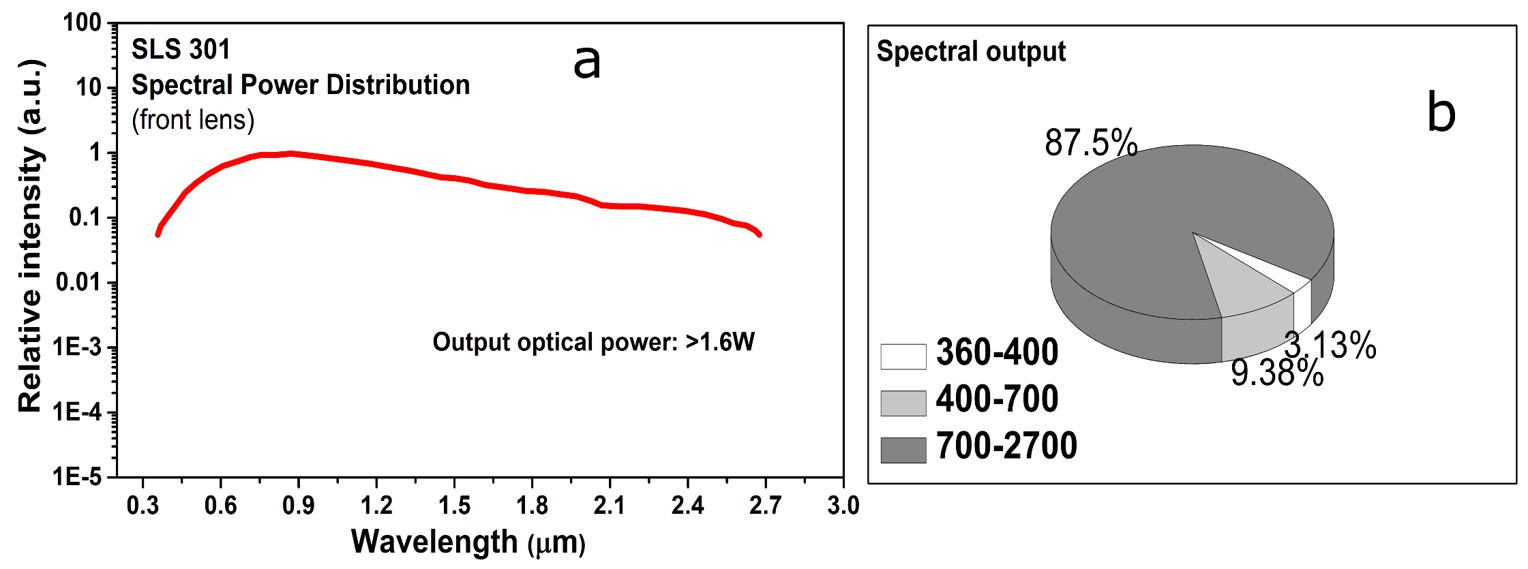}
		\caption{(a) Spectral power distribution of ThorLabs Quartz-Tungsten Lamp, SLS 301 (b) typical optical output power.}
		\label{fig:S4a}
	\end{center}
\end{figure}

The illumination to study the photo-response was from a Quartz-Tungsten lamp [see Fig. \ref{fig:S4a} for the lamp spectra] constricted to $\sim$1.5mm spot size (output optical power$\sim$ 460mW/cm$^2$).

\section{Electrical Measurements}
Pseudo four-probe contacts were made with silver(Ag) paint (Pelco Ted Pella, Inc.) with an active device area of 0.21 $\pm$ 0.03 $cm^2$ for conducting the electrical (resistance and current-voltage[$IV$] characteristics) and photo-response [PR] measurements. All electrical measurements were performed using a Keithley 2400 sourcemeter. The $IV$ characteristics were recorded by sourcing current and measuring voltage across the devices.

\subsection{Opto-electrical measurements}
 Photoresponse ($PR$) measurements were carried out by illuminating samples with a Stabilized Quartz-Tungsten Lamp(Thorlabs SLS301), with band-pass, neutral density (ND), $\lambda$  and long pass (LP) filters added for varying intensity and to record spectral dependence of $PR$. The optical power of illuminations are measured using a Thorlabs powermeter(PM100D) and photodiode power sensors(S122C \& S120VC). 

\begin{figure}[H]
	\begin{center}
		\includegraphics[width=0.75\textwidth]{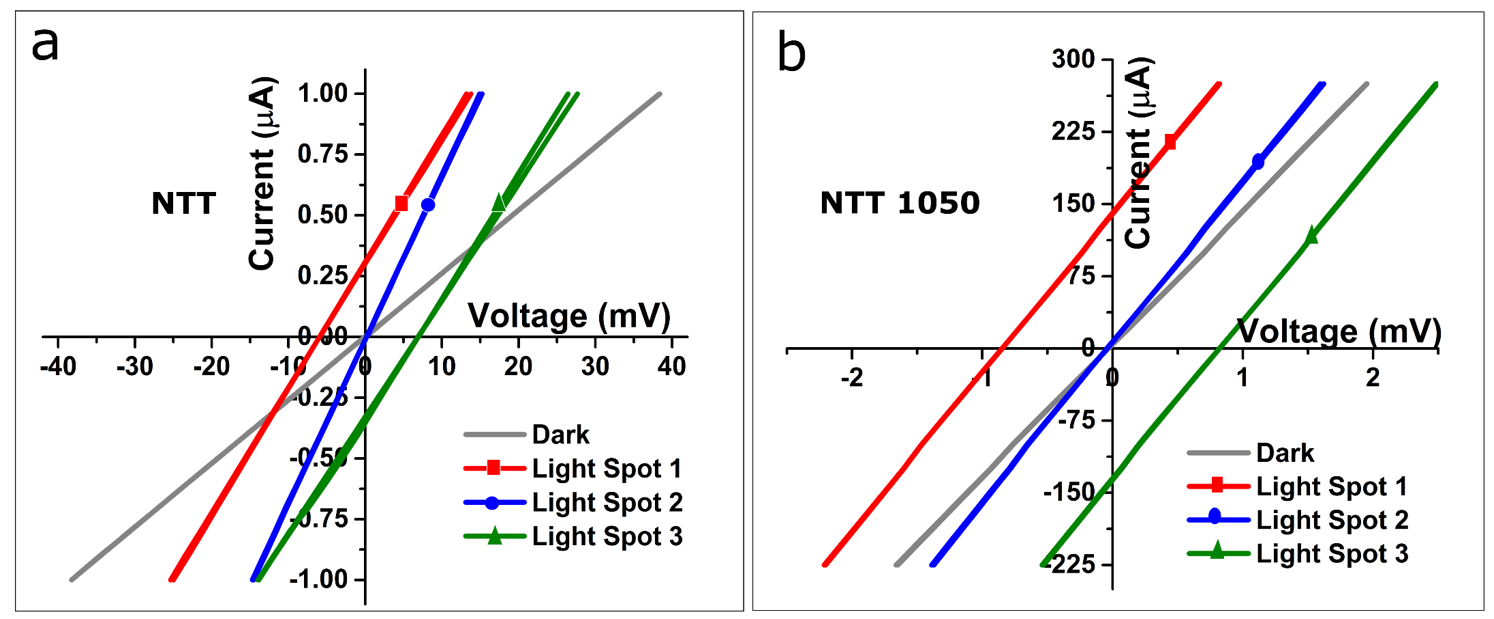}
		\caption{$IV$ plots in dark as well as with different illumination spots (white light) for (a) NTT \& (b) NTT1050.}
		\label{fig:S6}
	\end{center}
\end{figure}

\begin{figure}[H]
	\begin{center}
		\includegraphics[width=0.75\textwidth]{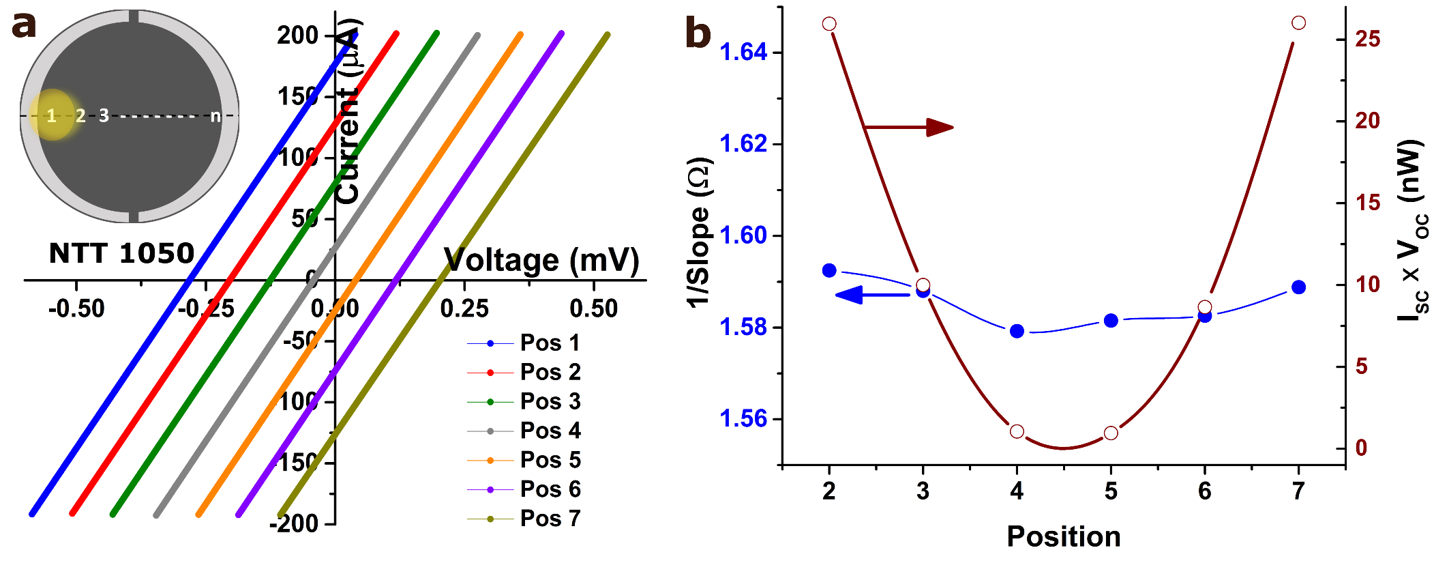}
		\caption{(a) $IV$s while scanning the focused white light spot across the NTT1050 and (b) the variation in resistance and power generated. The IVs shift as the illumination spot is scanned across the surface of NTT1050. While the resistance (1/slope, estimating the photoconductance) of the system is not varying much with the scanning, a significant trend is found in the ISC x VOC (estimate of photovoltaic effect) showing maximum response with (asymmetric illumination).}
		\label{fig:S6a}
	\end{center}
\end{figure} 

\subsubsection{Spectral resolved response}
\begin{figure}[H]
	\begin{center}
		\includegraphics[width=0.45\textwidth]{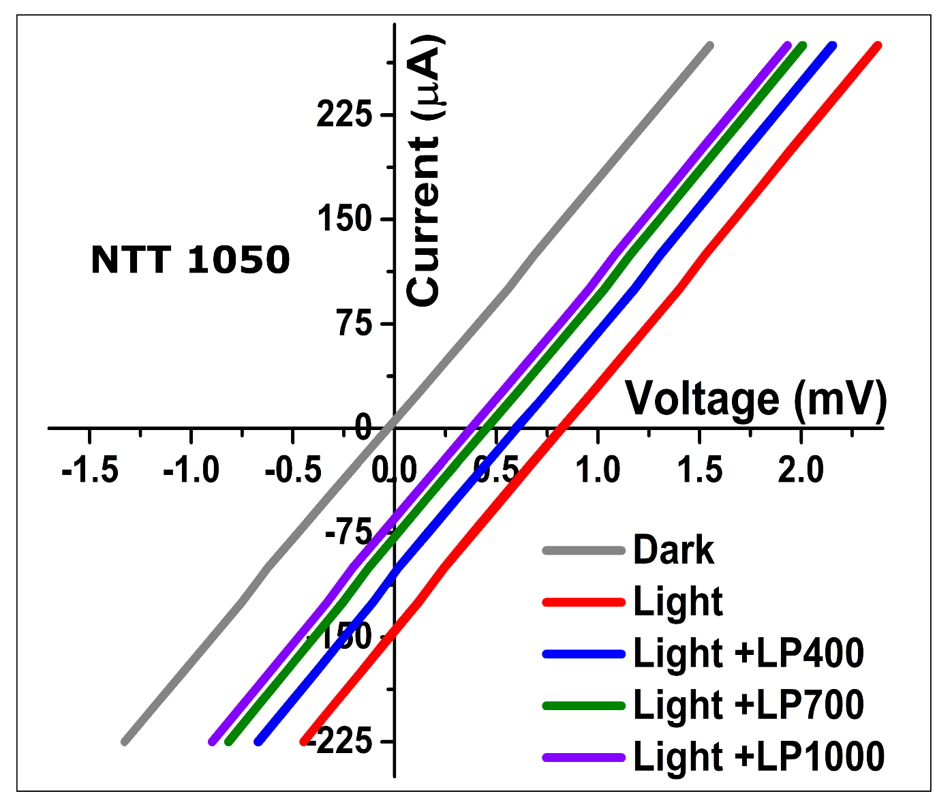}
		\caption{$IV$ plots of NTT1050 performed with illumination using different long pass (LP) filters on white light. (LP400, LP700, LP1000 - filters with cut-on wavelength at 400, 700 and 1000 nm, respectively.)}
		\label{fig:S6b}
	\end{center}
\end{figure}

\begin{figure}[H]
	\begin{center}
		\includegraphics[width=0.9\textwidth]{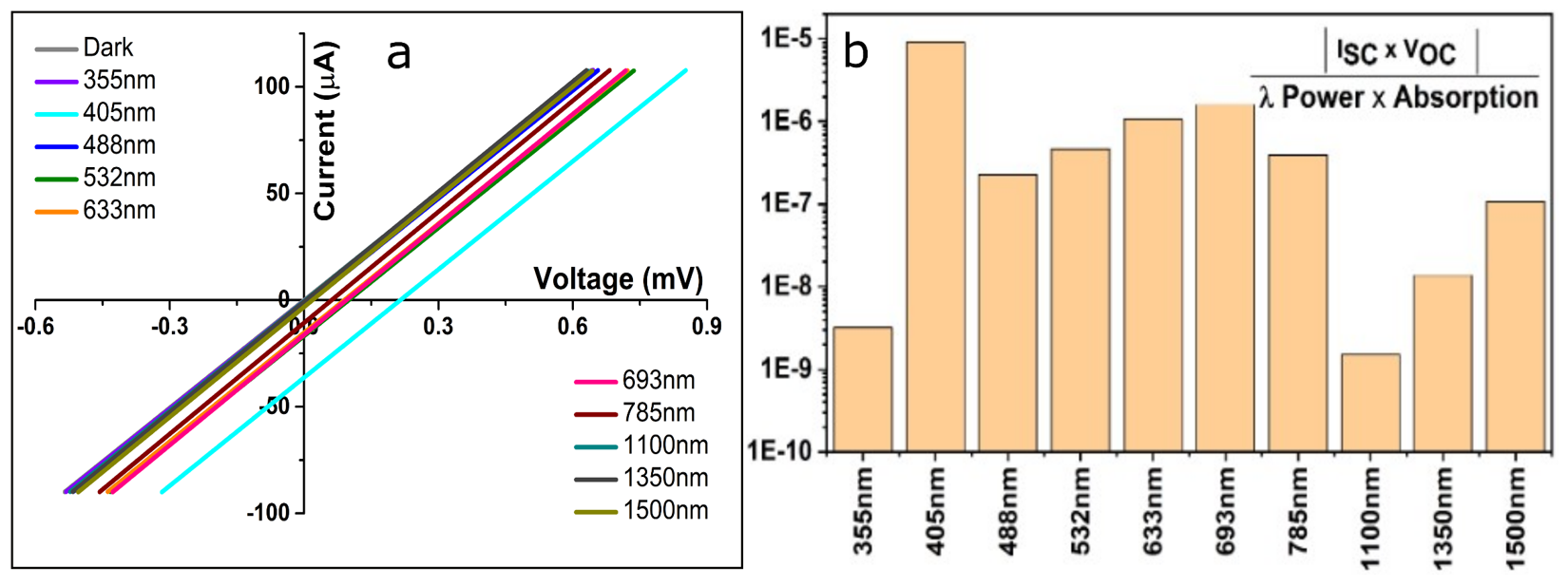}
		\caption{(a) $IV$ plots  of NTT1050 performed with different wavelength illuminations using wavelength selective filters with white light. (b) Photovoltaic power scaled with the measured input optical power of illumination and the absorption of NTT1050 at the corresponding wavelengths.}
		\label{fig:S7}
	\end{center}
\end{figure}

\begin{figure}[H]
	\begin{center}
		\includegraphics[width=0.8\textwidth]{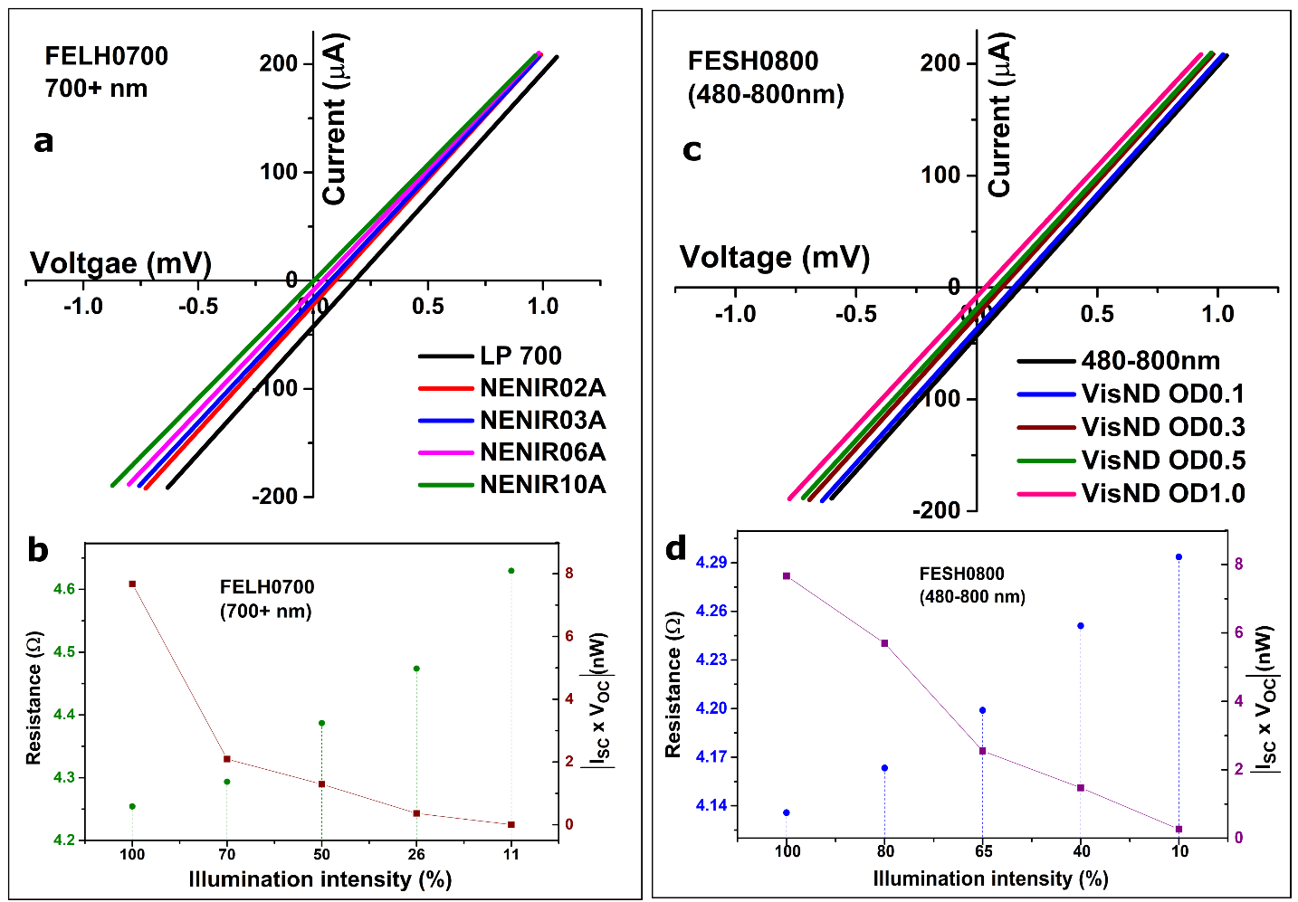}
		\caption{Photo-response of NTT1050 in (a) above 700nm and (c) 480-800nm range of the lamp spectrum with respective neutral density filters. (b \& d) Variation of resistance and photovoltaic power with the illumination intensity in the corresponding spectral ranges.}
		\label{fig:S8}
	\end{center}
\end{figure}

\subsection{Thermo-voltage measurements}
The thermo-voltaic measurements were done using Keysight DAQ6510 in a custom made probe station. The samples are fixed using Ag paint to two copper blocks kept side by side, where one of the copper blocks is equipped with a catridge heater. Thermocouples and electrical probes are attached to the sample sides to monitor the change in voltage with temperature ($\Delta$ V \& $\Delta$ T).
\begin{figure}[H]
	\begin{center}
		\includegraphics[width=0.95\textwidth]{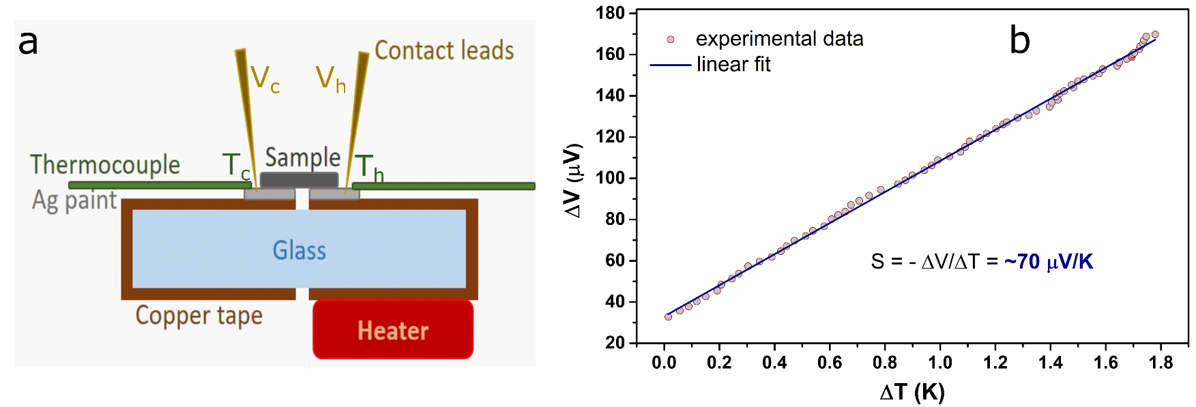}
		\caption{(a) Scheme of a custom made thermo-voltage measurement set-up (b) data and (c) the $\Delta$V vs $\Delta$T plot of the cooling region.}
		\label{fig:S5}
	\end{center}
\end{figure}
\subsubsection{Comparison of photo- \& thermo-voltages}
\begin{figure}[H]
	\begin{center}
		\includegraphics[width=0.95\textwidth]{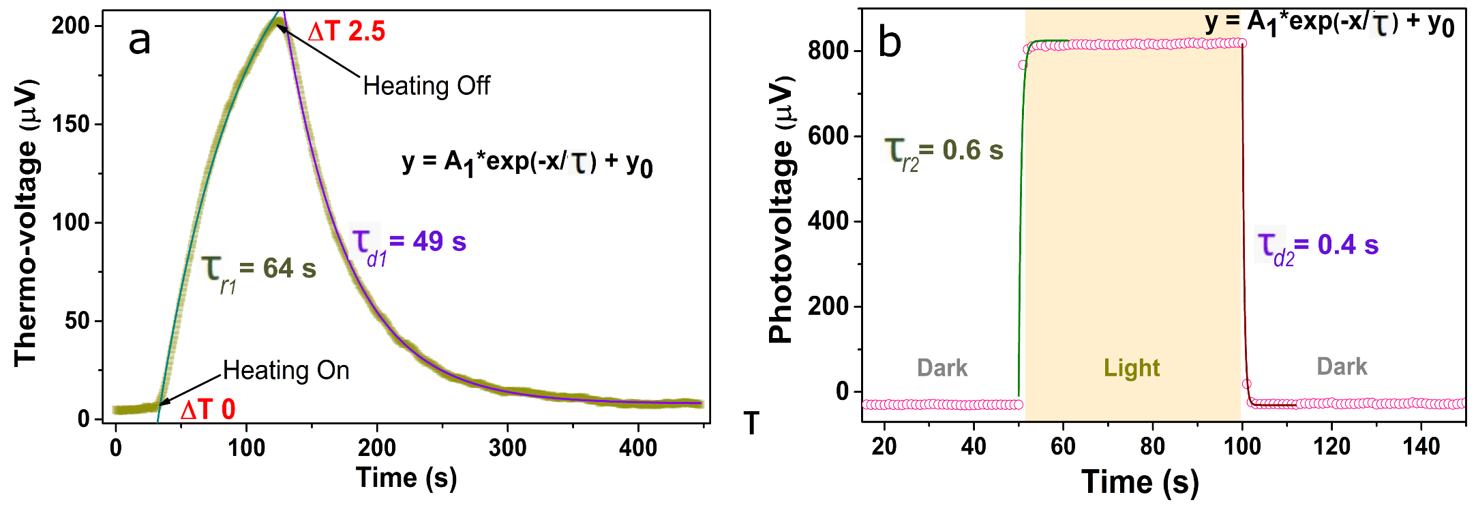}
		\caption{Comparison between temporal variation of (a) thermo-voltage and (b) photovoltage in NTT1050. Exponential fits of corresponding data provides the rise time ($\tau$$_r$) and decay time ($\tau$$_d$) constants.}
		\label{fig:S5a}
	\end{center}
\end{figure}

\subsection{Temperature dependent measurements}
All temperature dependent electrical measurements were performed with the sample mounted on a closed cycle cryostat (ARS Inc.) with GE varnish, connected to a Lake Shore 336 temperature controller.

\subsubsection{Effect of thermal sink}
Thermal maps were generated using Fluke Ti480 PRO Infrared Camera.
\begin{figure}[H]
	\begin{center}
		\includegraphics[width=0.75\textwidth]{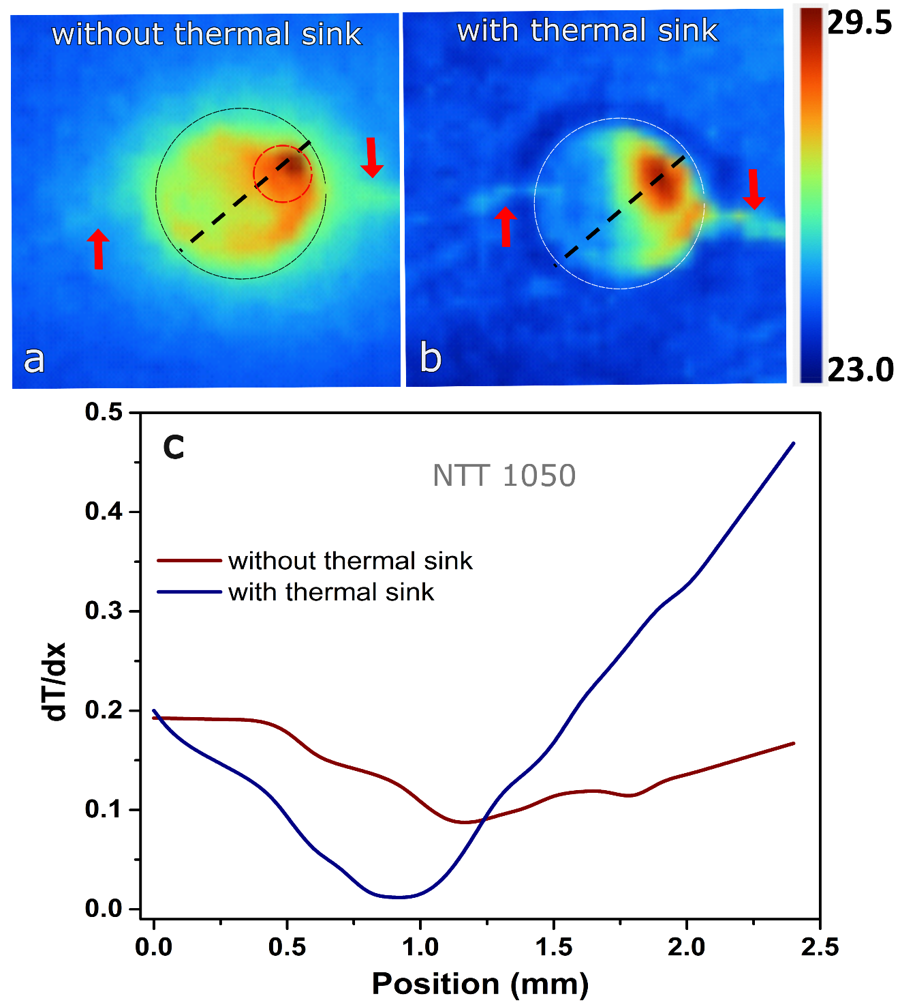}
		\caption{Thermal imaging of NTT 1050 (a) without thermal sink and (b) with thermal sink. (c) The variation of dT/dX from the heat maps along the dotted line in (a) and (b). Red circle in (a) shows the spot of illumination, and red arrows shows the electrical contact wires. The thermal imaging studies shows the role of thermal sink in causing higher thermal gradient and thereby stronger thermal current (I$_{th}$).}
		\label{fig:S9}
	\end{center}
\end{figure}

\subsubsection{Low temperature measurements }

\begin{figure}[H]
	\begin{center}
		\includegraphics[width=0.8\textwidth]{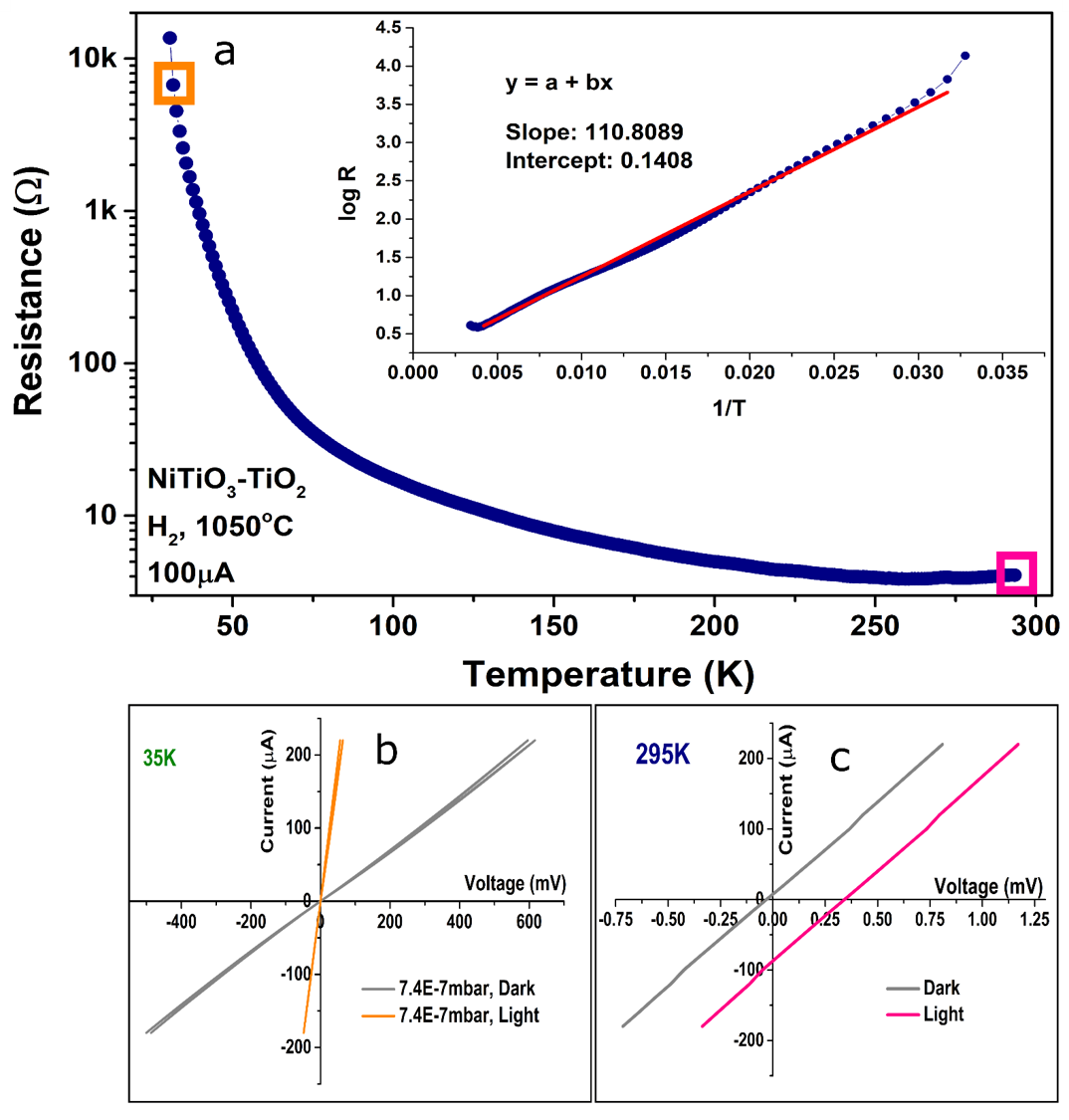}
		\caption{(a) $R vs T$ plot of NTT1050 with $logR$ $vs$ $\frac{1}{T}$ plot in the inset. A linear fit of the data provides E$_A$ $\sim$ 10 meV as E$_A$/k = slope from R(T) $\propto$ exp(E$_A$/kT). 
			 Bottom panel shows the $IV$ plots in dark as well as with illumination in (b) low temperature and at (c) room temperature.}
		\label{fig:S9a}
	\end{center}
\end{figure}

\begin{figure}[H]
	\begin{center}
		\includegraphics[width=0.65\textwidth]{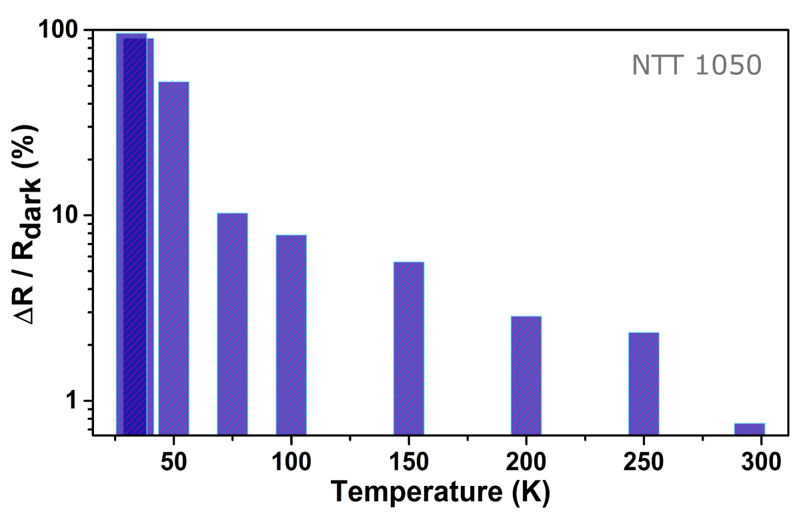}
		\caption{Variation of photo-conductance with temperature for NTT1050.}
		\label{fig:S10}
	\end{center}
\end{figure}

\section{Mechanism}
\subsection{Energy Band diagrams}
\begin{figure}[H]
	\begin{center}
		\includegraphics[width=0.45\textwidth]{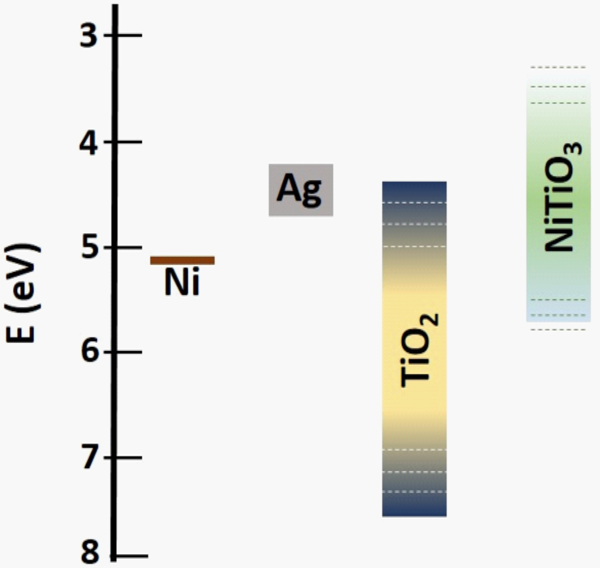}
		\caption{Energy Band diagrams of materials used in this study. \cite{michaelson1977work,yi2021heterostructure,yip2012recent} The work functions of Ni and Ag are presented along with the HOMO-LUMO (E$_V$ - E$_C$) levels of semiconducting \ch{TiO2} and \ch{NiTiO3} from the literature. The dotted lines signify the variation or spread of energy levels with different phases and defect levels from the literature survey.}
		\label{fig:S8a}
	\end{center}
\end{figure}

\begin{figure}[H]
	\begin{center}
		\includegraphics[width=0.5\textwidth]{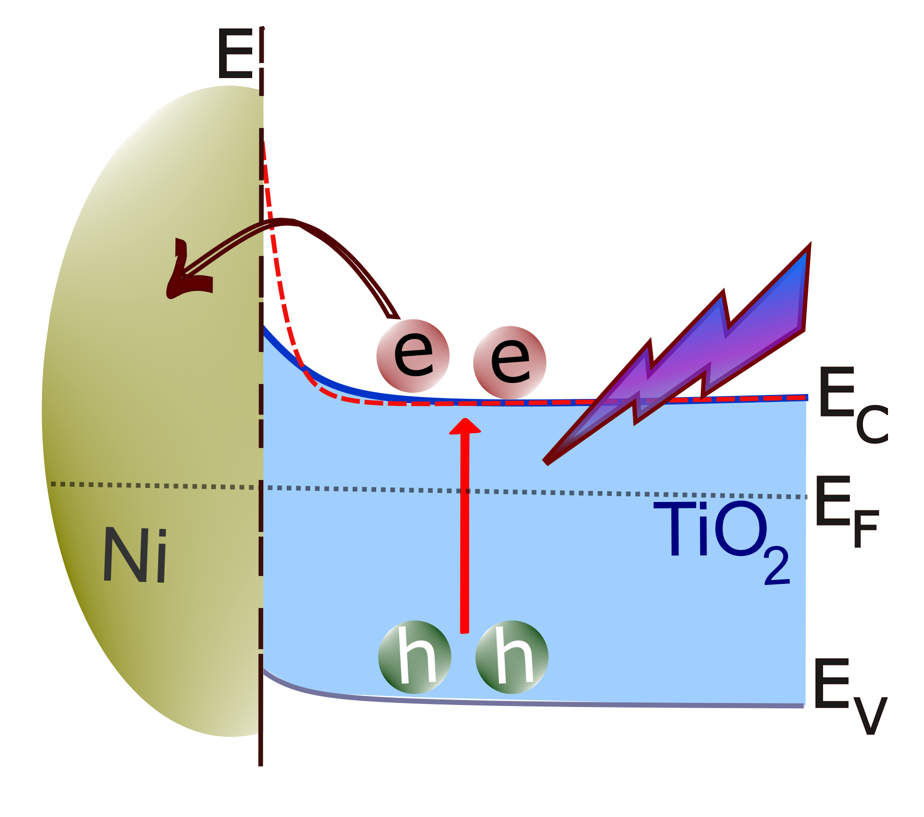}
		\caption{Mechanism of charge transfer between Ni NPs, Ag and \ch{TiO2} with illumination. Red dotted line shows the Conduction Band (CB) bending in the dark condition, and the solid blue line shows the CB under illumination. Ni-\ch{TiO2} forms a schottky junction in the dark of which the built-in potential significantly drops under illumination enabling the photo-generated carrier (electron) transfer from \ch{TiO2} to Ni. As this case is presented with stoichiometric \ch{TiO2}, it is expected to be similar in response with black-\ch{TiO2}, sub-oxides and \ch{mTiO2} as they will have smaller band-gaps.} 
		\label{fig:S8b}
	\end{center}
\end{figure}

\subsection{Finite element modelling of Ni-\ch{TiO2} schottky junction}
\begin{figure}[H]
	\begin{center}
		\includegraphics[width=0.95\textwidth]{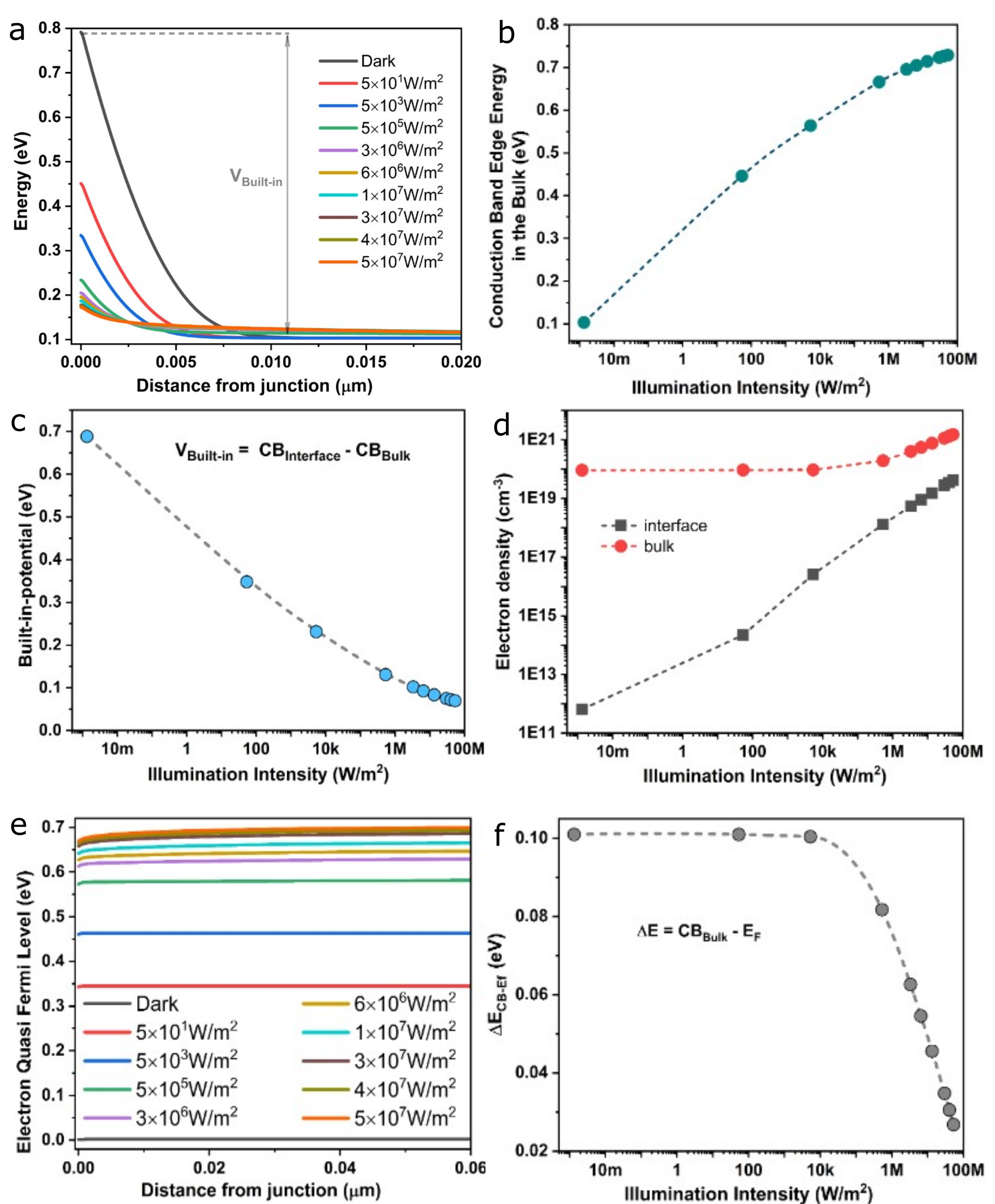}
		\caption{Finite element modelling results. (a) Evolution of Conduction Band (CB) of Ni-\ch{TiO2} junction as a function of illumination intensity, offset to the dark CB bulk region. (b) CB edge in the bulk of \ch{TiO2} with the corresponding illumination intensity, which is presented with offset in (a). (c) The built-in-potential at the Ni-\ch{TiO2} interface estimated as the difference of CB energy in the interface to the bulk (as depicted in (a)). (d) Change in electron density in the interface and bulk region with different illumination intensities. (e) Evolution of electron quasi fermi level (E$_F$) with illumination and (f) the energy difference between CB and E$_F$ to various illumination intensities.}
		\label{fig:S11}
	\end{center}
\end{figure}

Finite element modelling of a Schottky junction between Ni and \ch{TiO2} was carried out using COMSOL 6.0 in the semiconductor module to observe and analyse the band characteristics and electron distribution in such a system. 
The material properties of \ch{TiO2}  and Ni as well as other parameters are mentioned in table \ref{tab1}. 
First a bare \ch{TiO2}  sample is simulated without a Ni contact and to calibrate the system and a Ni contact was added at one end of the system model and the junction property was defined as Schottky type. The doping is defined as a donor type with a energy of the dopant state lying 20 meV below the conduction band.  Optical transitions were added to the simulation to observe the effect of electromagnetic (EM) waves on such a system. 
\begin{table}[H]
\caption{Parameters used in COMSOL Simulation using semiconductor module}
\centering
\begin{tabular}{|l|l|l|}
	\hline
	Item&Parameters&Value\\
	\hline
    \multirow{1}{4em}{\ch{TiO2}}&Relative permittivity&86\\
	&Bandgap&3.2 eV\\
	&Electron affinity&4.4 eV\\
	&Effective DOS, Valence Band&$4\times 10^{21}$ $cm^{-3}$\\
	&Effective DOS, Conduction Band&$4\times 10^{21}$ $cm^{-3}$\\
	&Electron mobility&0.1 $cm^{2} V^{-1}s^{-1}$\\
	&Hole mobility&0.025 $cm^{2} V^{-1}s^{-1}$\\
	&Refractive index, Real part&2.8736\\
	&Refractive index, Imaginary part&0\\
	&Doping density&1×$10^{20}$ $cm^{-3}$\\
	&Donar level&20 meV below CB\\
	\hline
	\multirow{1}{4em}{\ch{Ni}}&Work function&5.1 eV\\
	&Bias Voltage&0 V\\
	\hline
	\multirow{1}{4em}{Geometry}&1D&\\
	&\ch{TiO2}&Length 1 $\mu$m\\
	&Ni&Point contact\\
	&Meshing resolution&0.4 nm\\
	\hline
	Temperature&297 K&\\
	\hline
	\multirow{1}{4em}{Optical Transitions}&Wavelength&371 nm\\
	&Spontaneous life-time of carriers&2 ns\\
	&Intensity&$5\times 10^1$ to $5\times 10^7$ Wm$^{-2}$\\
	& & ($2\times 10^2$ to $2\times 10^5$ Vm$^{-1}$)\\
	\hline
	\multirow{1}{4em}{Tolerance}&Dark&1$\times$ 10$^{-4}$\\
	&Illumination&1$\times$ 10$^{-6}$\\
	\hline
\end{tabular}
\label{tab1}
\end{table}
 The results of the finite element modeling of Ni-\ch{TiO2} schottky junction is presented in figure \ref{fig:S11}. The conduction band (CB) edge of \ch{TiO2} (\ch{mTiO2}) systematically shifts up with illumination as presented in fig. \ref{fig:S11}b and the evolution of CB profile, offset to the dark level, is presented in fig. \ref{fig:S11}a. The schottky nature of the contact is evident and the built-in-potential, estimated by the energy difference of the band bending (CB$_{Interface}$-CB$_{Bulk}$), reduces with the increasing intensity of illumination as shown in fig. \ref{fig:S11}c. It can be discerned that the increase in illumination intensity methodically changes the electron density (Fig. \ref{fig:S11} d) and the fermi level moves closer to the CB (Figs. \ref{fig:S11}e \& f) causing the reduced band bending. The reduced built-in-potential implies an easier cross-over for the photo-generated electrons and thus enabling efficient charge segregation and diffusion to the Ag contact.

\bibliography{bibfile.bib}
	
\end{document}